\documentclass[sigconf]{acmart}
\copyrightyear{2024} 
\acmYear{2024} 
\usepackage{bbm}
\usepackage{balance}
\usepackage{cleveref}
\usepackage{multirow}
\usepackage{graphicx}
\usepackage{enumitem}
\usepackage{xcolor}
\usepackage{makecell} 
\usepackage{colortbl}
\usepackage{algorithm,algpseudocode}

\usepackage{bm}
\usepackage{color}
\usepackage{apxproof}
\usepackage{gensymb}
\usepackage{pifont}%

\usepackage{xspace}
\usepackage{mathtools}
\usepackage{subcaption}

\newcommand{\squishlist}{
 \begin{list}{$\bullet$}
  { \setlength{\itemsep}{0pt}
     \setlength{\parsep}{3pt}
     \setlength{\topsep}{3pt}
     \setlength{\partopsep}{0pt}
     \setlength{\leftmargin}{1.5em}
     \setlength{\labelwidth}{1em}
     \setlength{\labelsep}{0.5em} } }

\newcommand{\squishlisttwo}{
 \begin{list}{$\bullet$}
  { \setlength{\itemsep}{0pt}
     \setlength{\parsep}{0pt}
    \setlength{\topsep}{0pt}
    \setlength{\partopsep}{0pt}
\setlength{\leftmargin}{2em}
\setlength{\labelwidth}{1.5em}
\setlength{\labelsep}{0.5em} } }

\newcommand{\squishend}{
\end{list}  }

\AtBeginDocument{%
  \providecommand\BibTeX{{%
    \normalfont B\kern-0.5em{\scshape i\kern-0.25em b}\kern-0.8em\TeX}}}

\begin{document}

\title{TiM4Rec: An Efficient Sequential Recommendation Model Based on Time-Aware Structured State Space Duality Model}

\author{Hao Fan~$\dag$}
\affiliation{
  \institution{Zhejiang A\&F University}
  \city{HangZhou}
  \country{China}}
\email{fanhao986486@stu.zafu.edu.cn}

\author{Mengyi Zhu~$\dag$}
\affiliation{
  \institution{Soochow University}
  \city{SuZhou}
  \country{China}}
\email{myzhu1@stu.suda.edu.cn}

\author{Yanrong Hu~$\dag$*}
\affiliation{
  \institution{Zhejiang A\&F University}
  \city{HangZhou}
  \country{China}}
\email{yanrong_hu@zafu.edu.cn}

\author{Hailin Feng~}
\affiliation{
  \institution{Zhejiang A\&F University}
  \city{HangZhou}
  \country{China}}
\email{Hlfeng@zafu.edu.cn}

\author{Zhijie He~}
\affiliation{
  \institution{Zhejiang A\&F University}
  \city{HangZhou}
  \country{China}}
\email{2023111021004@stu.zafu.edu.cn}

\author{Hongjiu Liu~}
\affiliation{
  \institution{Zhejiang A\&F University}
  \city{HangZhou}
  \country{China}}
\email{joe_hunter@zafu.edu.cn}

\author{Qingyang Liu~}
\affiliation{
  \institution{Georg-August-Universität Göttingen}
  \city{Göttingen}
  \country{Germany}}
\email{qingyang.liu@stud.uni-goettingen.de}

\renewcommand{\shortauthors}{Fan, et al.}
\thanks{* ~ The corresponding author. \\
$\dag$~The authors contribute equally to this paper. \\
© [2024] [Hao Fan~$\dag$, Mengyi Zhu~$\dag$, Yanrong Hu~$\dag$*, Hailin Feng, Zhijie He, Hongjiu Liu, and Qingyang Liu.] All rights reserved. This paper may be used for personal or academic purposes provided proper citation is given. Reproduction, distribution, or commercial use is prohibited without written permission from the authors.}

\begin{abstract}
  The Sequential Recommendation modeling paradigm is shifting from Transformer to Mamba architecture, which comprises two generations: Mamba1, based on the State Space Model (SSM), and Mamba2, based on State Space Duality (SSD). Although SSD offers superior computational efficiency compared to SSM, it suffers performance degradation in sequential recommendation tasks, especially in low-dimensional scenarios that are critical for these tasks. Considering that time-aware enhancement methods are commonly employed to mitigate performance loss, our analysis reveals that the performance decline of SSD can similarly be fundamentally compensated by leveraging mechanisms in time-aware methods. Thus, we propose integrating time-awareness into the SSD framework to address these performance issues. However, integrating current time-aware methods, modeled after TiSASRec, into SSD faces the following challenges: 1) the complexity of integrating these transformer-based mechanisms with the SSD architecture, and 2) the computational inefficiency caused by the need for dimensionality expansion of time-difference modeling. To overcome these challenges, we introduce a novel Time-aware Structured Masked Matrix that efficiently incorporates time-aware capabilities into SSD. Building on this, we propose \textbf{Ti}me-Aware \textbf{M}amba for \textbf{Rec}ommendation (TiM4Rec), which mitigates performance degradation in low-dimensional SSD contexts while preserving computational efficiency. This marks the inaugural application of a time-aware enhancement method specifically tailored for the Mamba architecture within the domain of sequential recommendation. Extensive experiments conducted on three real-world datasets demonstrate the superiority of our approach. The code for our model is accessible at \href{https://github.com/AlwaysFHao/TiM4Rec}{https://github.com/AlwaysFHao/TiM4Rec}.
\end{abstract}
\settopmatter{printfolios=true}

\maketitle
\thispagestyle{empty}
\section{Introduction}
Sequential Recommendation (SR) is an important branch of Recommendation Systems (RS), leveraging the user's historical interaction sequence to predict the items that the user may interact with next \cite{RSSurvey2005, SRSurvey2018}. Unlike traditional methods—such as collaborative filtering \cite{CollaborativeFiltering2001} and content-based filtering \cite{VBPR2016}—that predominantly depend on static user features and past behaviors, SR focuses on capturing dynamic user behavior patterns, making it more effective in adapting to rapid shifts in user interests and diverse consumption habits \cite{Deezer2022, SASRec2018, RSSurvey2005}.
\par
The landscape of SR models has evolved significantly, transitioning from early models based on Markov chains \cite{HRM2016, FPMC2010} to those leveraging Convolutional Neural Networks (CNNs) \cite{TangW2018}, Recurrent Neural Networks (RNNs) \cite{GRU4Rec2016, LiRCRLM2017}, Graph Neural Networks (GNNs) \cite{ChangGZHNSJ02021, SRGNN2019}, notable examples include Caser \cite{TangW2018} and GRU4Rec \cite{GRU4Rec2016}. 
However, these models often suffer from performance limitations inherent to their core architectures, resulting in suboptimal prediction accuracy. 
The introduction of the Transformer architecture \cite{Transformer2017}, grounded in an attention mechanism, has marked a paradigm shift due to its superior performance and parallelizable training capabilities and made it a mainstream framework across various domains that have also significantly influenced the sequential recommendation domain \cite{BTMT2024, SASRec2018, BERT4Rec2019, AFMN2022}. Although Transformer-based sequential recommendation models, such as SASRec \cite{SASRec2018}, have demonstrated promising performance, their inherent quadratic computational load of the Transformer architecture poses significant challenges for efficient sequential recommendation modeling \cite{RecBLR2024, Mamba4Rec2024, LRURec2024}.
% In recent years, as evidenced by models like SASRec \cite{SASRec2018} and BERT4Rec \cite{BERT4Rec2019} effectively adapted the success of Transformers to this area. 
\par
To mitigate the Transformer’s inherent computational burden, numerous models with linear complexity have been proposed (e.g., HiPPO \cite{HiPPO2020}, S4 \cite{S4-2022}, LRU \cite{LRU2023} and RWKV \cite{RWKV2023}). Among these, Mamba \cite{Mamba2023} strikes a commendable balance between performance and efficiency. Mamba is derived from the State Space Model (SSM), building upon prior works (e.g., HiPPO \cite{HiPPO2020}, S4 \cite{S4-2022}, and S5 \cite{S5-2023}), and has been widely and successfully applied across various domains \cite{MambaNLP2024, VisionMamba2024, SSMEC2024}. For efficient sequential recommendation modeling, Mamba4Rec \cite{Mamba4Rec2024} is the first to adapt the Mamba architecture to SR. Experimental results indicate that the Mamba architecture holds greater potential than the Transformer in SR tasks \cite{RecBLR2024, Mamba4Rec2024}. Subsequently, Dao et al. \cite{SSD2024} optimized the architecture from a parallel computing perspective, proposing a novel architecture called State Space Duality (SSD) based on Mamba, also referred to as Mamba2. SSD demonstrates superior efficiency compared to SSM and exhibits enhanced performance in high-dimensional modeling \cite{SSD2024}; however, its performance in low-dimensional settings is suboptimal (further analysis will be provided later). Although SSD4Rec \cite{SSD4Rec2024} achieves better performance in higher dimensions than Mamba4Rec through techniques such as variable-length sequence training and bidirectional SSD, it does not outperform Mamba4Rec in lower dimensions. Pure ID sequential recommendation typically achieves optimal performance in low dimensions \cite{BERT2019, GRU4Rec2016, SASRec2018, LRURec2024}, and high-dimensional modeling contradicts the objective of efficient sequential recommendation. If we solely leverage the advantages of the Mamba2 architecture for high-dimensional modeling in SR tasks, without considering the applicability of high-dimensional processing in pure ID-based SR scenarios, the associated computational cost may outweigh its benefits. \textbf{Therefore, addressing the performance deficit of the SSD architecture in low dimensions becomes a crucial challenge for efficient sequential recommendation modeling.}
\par
\begin{figure}[t]
    \centering
    \includegraphics[width=0.45\textwidth]{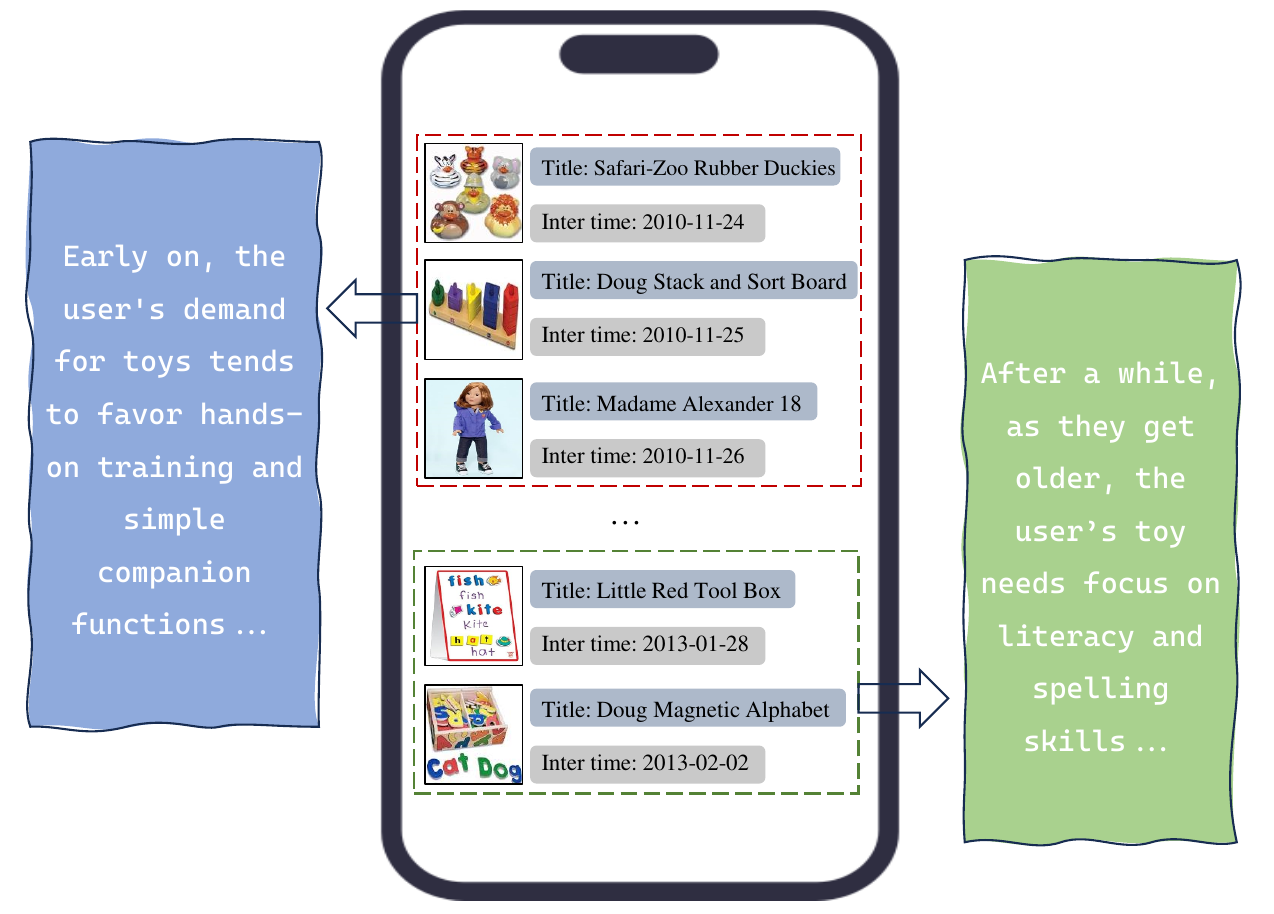}
    \caption{User interest transfer. Over time, a noticeable shift in the user's points of interest is observed, reflecting a common characteristic of user interaction behaviors.}
    \label{fig:interest-transfer}
\end{figure}
User interaction behaviors typically exhibit clear temporal patterns \cite{TiCoSeRec2023, TiSASRec2020}. As illustrated in Figure \ref{fig:interest-transfer}, we extract an interaction sequence from a real-world dataset \cite{Amazon2014}, clearly demonstrating that user interests tend to change systematically over time. Previous Transformer-based SR models have not fully leveraged these temporal dynamics \cite{MusicRec2020, TiSASRec2020, MOJITO2023}. Consequently, numerous time-aware enhancement models have emerged \cite{BTMT2024, TimelyRec2021, MEANTIME2020, TiSASRec2020, MOJITO2023, TAT4SRec2023} based on the TiSASRec \cite{TiSASRec2020} modeling paradigm, aiming to improve the recommendation performance of Transformers by effectively incorporating time-awareness.
By analyzing the reasons for the performance degradation of SSD compared to SSM, we find that, theoretically, time-aware enhancement methods can address the performance loss issue of SSD in low-dimension. Therefore, we propose integrating time-aware enhancement methods into the SSD architecture to tackle this problem. However, the existing time-aware augmentation methods encounter the following issues when integrated to SSD architectures:
\par
\textbf{Challenge \uppercase\expandafter{\romannumeral1}}: \textit{Incompatibility with the SSD architecture.} Current time-aware enhancement methods based on Transformers cannot be effectively integrated into the computational core of SSD. Specifically, time-aware enhancement methods following the TiSASRec paradigm fundamentally apply time-difference information to the attention score matrix. However, since SSD lacks an explicit representation of the attention score matrix during its computation, existing time-aware enhancement methods are incompatible with SSD's architecture.
\par
\textbf{Challenge \uppercase\expandafter{\romannumeral2}}: \textit{Low computational efficiency.} In the paradigm of time-aware enhancement methods based on TiSASRec, it is necessary to increase the dimensionality of the time difference matrix to obtain a time-awareness score matrix, which is then integrated into the attention score matrix. This approach results in significant additional computational resource consumption, rendering existing time-aware enhancement methods ineffective for efficient sequential recommendation tasks.
\par
To address the aforementioned challenges, we introduce, for the first time, a time-aware enhancement method specifically tailored for SSD architectures that maintain computational efficiency while mitigating performance degradation issues. Specifically, we propose the Time-aware Structured Masked Matrix, which integrates time-awareness into the SSD architecture. This method involves scalar-level transformations along the sequence dimension to convert time difference information into a structured mask matrix suitable for SSD architectures. Compared to traditional time-aware enhancement approaches, our method significantly improves computational efficiency.
It is noteworthy that the time-aware enhancement method proposed in this paper preserves the linear computational complexity advantage of the SSD architecture in the sequence length. 
Therefore, in comparison with the SASRec model \cite{SASRec2018}, which represents the Transformer architecture, and the Mamba4Rec model \cite{Mamba4Rec2024}, which represents the SSM architecture, it retains a significant advantage in terms of model training and inference speed. 
\par
We designate our work as \textbf{TiM4Rec}(\textbf{Ti}me-aware \textbf{M}amba for \textbf{Rec}ommendation) and outline its primary contributions as follows:
\begin{itemize}[leftmargin=10pt]
    \item To the best of our knowledge, we conduct a pioneering exploration of a time-aware enhancement method tailored for the linear computational complexity of the Mamba architecture in the field of sequential recommendation.
    \item Our theoretical analysis reveals the underlying reasons for the performance degradation of SSD compared to SSM. In response, we propose a targeted time-aware enhancement method designed to mitigate the performance loss of SSD in sequential recommendation tasks.
    \item We integrate time-awareness into the SSD architecture by constructing a Time-aware Structured Masked Matrix. In contrast to the time-aware enhancement method modeled on TiSASRec, which requires a computational complexity of $O(T^2 N)$, our approach achieves efficient time-awareness with only $O(T)$ computational complexity.
    \item Experiments on three real-world public datasets demonstrate the superiority of our proposed TiM4Rec compared with Transformer-based and SSM-based baselines in terms of both effectiveness and efficiency. 
\end{itemize}

\section{RELATED WORK}
    \label{sec:pre}
    \subsection{Sequential Recommendation}
        Previous research on sequential recommendation systems focused primarily on utilizing Recurrent Neural Networks (RNNs) \cite{GRU4Rec2016, LiRCRLM2017}. However, due to the performance limitations of RNNs and their inefficiency in parallel computation, an increasing number of successful studies in sequential recommendation have shifted towards employing Transformer-based approaches \cite{BTMT2024, SASRec2018, BERT4Rec2019, AFMN2022}. The most representative model is SASRec \cite{SASRec2018}, which captures user interest features from user interaction sequences through the causal self-attention mechanism \cite{Transformer2017}. Owing to the powerful modeling capability of self-attention, SASRec achieves strong recommendation performance. However, the Transformer architecture requires a quadratic computational load with respect to the sequence length, making it inefficient for sequence modeling (we will analyze the computational complexity in detail in Section \ref{sec:complexity}). 
        \par
        In the application of recommender systems, various platforms typically require real-time recommendations, as excessive latency can detrimentally impact user experience \cite{TencentRec2015, MaNLD2020}. Consequently, the investigation of efficient sequential recommendation methods has become imperative \cite{RecBLR2024, LRURec2024}. Recently, many models with linear computational loads have emerged as alternatives to the original Transformer architecture \cite{HiPPO2020, S4-2022, LRU2023, RWKV2023, S5-2023}, among which Mamba \cite{Mamba2023} has achieved notable success and has been promoted in various other fields \cite{MambaNLP2024, VisionMamba2024, SSMEC2024}. Mamba4Rec \cite{Mamba4Rec2024} is the first to adapt the Mamba architecture to the sequential recommendation domain. In essence, it employs a modeling method similar to SASRec, merely replacing the self-attention mechanism module in SASRec with the Mamba1 module. Since Mamba1 uses state transitions to finely control the flow of information within sequences, rather than the fully connected modeling approach of self-attention that considers each vector's contribution equally, it better aligns with user interaction habits (i.e., recent interactions more accurately reflect users' next interaction tendencies). Thus, Mamba4Rec achieves better performance while maintaining efficiency.
        \par
        With the introduction of the Mamba2 \cite{SSD2024} architecture, SSD4Rec \cite{SSD4Rec2024} has been applied to the sequential recommendation domain for the first time, further enhancing time efficiency compared to Mamba4Rec. Although SSD4Rec surpasses Mamba4Rec in high-dimensional modeling through additional processes such as variable-length sequence input and bidirectional modeling, this improvement is based on the premise that Mamba4Rec experiences a performance drop in high-dimensional settings (which we will verify in the experimental section). Prioritizing high-dimensional modeling contradicts the original intention of efficient modeling. For pure ID sequential recommendation, modeling is inherently more suitable in lower dimensions. 
    \subsection{Time-aware methods in SR}
        As mentioned in the introduction, user interaction behaviors typically exhibit clear temporal patterns \cite{TiCoSeRec2023, TiSASRec2020}. Consequently, researchers have introduced time-aware enhancement methods into the field of sequential recommendation. TiSASRec \cite{TiSASRec2020} represents the first investigation of a time-aware enhanced sequential recommendation algorithm based on the Transformer architecture. In essence, TiSASRec enhances the computation of the attention score matrix by constructing a time difference matrix and generating time difference positional encodings based on the difference values. MEANTIME \cite{MEANTIME2020} enhances the learning of various temporal encoding patterns by utilizing multiple attention heads. TimelyRec \cite{TimelyRec2021} models heterogeneous temporal patterns at different temporal granularities. MOJITO \cite{MOJITO2023} captures diverse fine-grained temporal patterns simultaneously by integrating an attention mechanism with Gaussian mixtures. TAT4SRec \cite{TAT4SRec2023} employs an encoder-decoder architecture to model timestamps and interaction items separately. Although these methods achieve improved performance compared to TiSASRec, they share the same modeling paradigm as TiSASRec and suffer from common limitations such as inefficiency and restriction to attention-based mechanisms.
        \par
        The methodological paradigm of TiSASRec, while capable of enhancing the performance of Transformers in sequential recommendation through the integration of time difference matrices, incurs a significant increase in computational overhead. This is due to the necessity of embedding the time difference matrices and executing attention computations before they can be incorporated into the final attention score matrix. Such an approach contradicts the principle of efficient computation. Therefore, discovering a method to directly integrate time difference matrices into the attention score matrix is crucial for applying time-aware enhancement techniques to sequential recommendation modeling efficiently.
        
    \subsection{State Space Models}
    \label{sec:ssm}
    The State Space Model(SSM) is a sequence modeling framework based on linear ordinary differential equations. 
    It maps the input sequence $x(t) \in \mathbb{R}^{D} $ to the output sequence $y(t) \in \mathbb{R}^{D}$ through the latent state $h(t) \in \mathbb{R}^N$:
    \begin{equation}
        \begin{aligned}
            h^{\prime}(t) & =\boldsymbol{A} h(t)+\boldsymbol{B} x(t), \\
            y(t) & =\boldsymbol{C} h(t) + \boldsymbol{D} x(t),
        \end{aligned}
        \label{eq:ssm}
    \end{equation}
    where $\boldsymbol{A} \in \mathbb{R}^{N \times N}$ and $\boldsymbol{B}, \boldsymbol{C} \in \mathbb{R}^{N \times D}$ are learnable matrices.
    To enable SSM to effectively represent discrete data, it is imperative to discretize the data in accordance with the given step size $\Delta$.
    An effective discretization method for Eq. \ref{eq:ssm} is the Zero Order Hold (ZOH) method \cite{SSM2021}.
    Assuming the previous time step is $t-1$ and the current time step is $t$, Eq. \ref{eq:ssm} can be solved by using the constant variation method:
    \begin{equation*}
        \begin{aligned}
            h(t) 
            & = \exp(\Delta \boldsymbol{A}) h(t - 1) + \int_{t - 1}^{t} \exp(\boldsymbol{A}\left(t - \tau\right)) \boldsymbol{B} x\left(\tau\right) d\tau \\
            & \approx \exp(\Delta\boldsymbol{A}) h(t - 1) + \int_{t - 1}^{t} \exp(\boldsymbol{A}\left(t - \tau\right)) d\tau \boldsymbol{B} x\left(t\right)\\
            % & = e^{\Delta A} h(t - 1) + A^{-1} \Delta^{-1} \left(e^{\Delta A} - I\right) \Delta B x\left(t\right) \\
            & = \exp(\Delta \boldsymbol{A}) h(t - 1) + \left(\Delta \boldsymbol{A}\right)^{-1} \left(\exp(\Delta \boldsymbol{A}) - I\right) \Delta \boldsymbol{B} x\left(t\right),
        \end{aligned}
    \end{equation*}
    We assume $\overline{\boldsymbol{A}}=\exp (\Delta \boldsymbol{A})$ and $\overline{\boldsymbol{B}}=(\Delta \boldsymbol{A})^{-1}(\exp (\Delta \boldsymbol{A})-\boldsymbol{I}) \cdot \Delta \boldsymbol{B}$, 
    then the following result can be derived:
    \begin{equation}
        \begin{aligned}
            h_t & =\overline{\boldsymbol{A}} h_{t-1}+\overline{\boldsymbol{B}} x_t, \\
            y_t & =\boldsymbol{C} h_t + \boldsymbol{D} x_t,
        \end{aligned}
    \end{equation}
    The Structured State Space Model (S4) \cite{S4-2022} is developed from the vanilla SSM. By imposing a structure on the state matrix $\boldsymbol{A} \in \mathbb{R}^{D \times N}$ through HiPPO \cite{HiPPO2020} initialization, 
    it further enhances the modeling of long-range dependencies.
    \par

    Based on the foundation established by S4 model, Mamba \cite{Mamba2023} advances the concept of selectivity by integrating parameter matrices $\Delta \in \mathbb{R}^{T \times D}$, $\boldsymbol{B} \in \mathbb{R}^{T \times N}$ and $\boldsymbol{C} \in \mathbb{R}^{T \times N}$, which vary over time. 
    Through these methodological advancements, Mamba is capable of selectively transforming the input sequence $\boldsymbol{X} \in \mathbb{R}^{T \times D}$ into the output sequence $\boldsymbol{Y} \in \mathbb{R}^{T \times D}$ as delineated by the following equation:
    \begin{equation}
        \begin{aligned}
            \boldsymbol{Y}_{i} &= \sum_{j} \left[\boldsymbol{C}_{i}^{\top} \left(\overline{\boldsymbol{A}}_{i} \cdot \overline{\boldsymbol{B}}_{j}\right)\right] \cdot X_{j}, \\
            \overline{\boldsymbol{A}}_{i} &= \prod_{k = 1}^{i - j} \exp(\Delta_{k}^{\top} \boldsymbol{A}), \\
            \overline{\boldsymbol{B}}_{j} &= \left(\Delta_{j}^{\top} \boldsymbol{A}\right)^{-1}\left(\exp(\Delta_{j}^{\top} \boldsymbol{A})-\boldsymbol{I}\right) \cdot \left(\Delta_{j}^{\top} \boldsymbol{B}_{j}\right),
        \end{aligned}
    \end{equation}
    Additionally, Mamba leverages CUDA to implement the Parallel Scan algorithm \cite{Parallelizing2018} and draws inspiration from FlashAttention \cite{FlashAttention2022} by employing hardware-aware parallel computation strategies to enhance computational efficiency.
    Through the aforementioned enhancements, Mamba demonstrates the capability to attain performance levels that are nearly on par with those of Transformers, while incurring only linear computational complexity with respect to the sequence length.
    \subsection{State Space Duality}
    \label{sec:ssd}
    The SSM framework, despite employing hardware-aware selective scan algorithms to facilitate parallel computing, 
    encounters limitations in effectively harnessing matrix computation units within the hardware. 
    This inefficacy stems from its non-reliance on matrix operations as a foundational approach. 
    Furthermore, Mamba's hardware-aware algorithm computational core resides entirely within the GPU's SRAM, 
    which is typically of limited capacity. 
    This restriction constrains the feature dimensionality in model computations, 
    thereby impeding the ability of the model to operate within higher-dimensional spaces.
    \par
    From a computational efficiency standpoint, 
    Dao et al. \cite{SSD2024} innovatively transformed matrix $\boldsymbol{A}$ 
    % (In this context, matrix $\boldsymbol{A}$ refers to the non-discretized matrix within the framework of SSM, as opposed to the discretized matrix $\overline{\boldsymbol{A}}$ )
    into a scalar value to facilitate SSM matrixization calculation and discovered that this variant of SSM exhibits duality with masked attention mechanisms. 
    To preserve the linear complexity benefit of SSM in the sequence dimension, 
    they leveraged the block properties of semi-separable matrices to implement linear attention computation, 
    culminating in what they termed the Structured State Space Duality.
    \par
    In essence, SSD, the structure they proposed, initially formulated SSM employing a matrix computation framework:
    \begin{equation}
            \boldsymbol{Y} = SSM(\boldsymbol{A}, \boldsymbol{B}, \boldsymbol{C})(X) = \boldsymbol{M}\boldsymbol{X}, 
    \end{equation}
    \begin{equation}
        \label{eq:ssm_m}
        \begin{aligned}
        &\boldsymbol{A}_{m:n}^{\times} \coloneqq \prod_{k=m}^{n}\boldsymbol{A}_{k} = \boldsymbol{A}_{n} \boldsymbol{A}_{n-1} \cdots \boldsymbol{A}_{m}, \\
        \boldsymbol{M} &= \begin{bmatrix}
            \boldsymbol{C}_{0}^{\top} \boldsymbol{B}_{0} & & & & \\
            \boldsymbol{C}_{1}^{\top} \boldsymbol{A}_{1} \boldsymbol{B}_{0} & \boldsymbol{C}_{1}^{\top} \boldsymbol{B}_{1} & & & \\
            \boldsymbol{C}_{2}^{\top} \boldsymbol{A}_{2} \boldsymbol{A}_{1} \boldsymbol{B}_{0} & \boldsymbol{C}_{2}^{\top} \boldsymbol{A}_{2} \boldsymbol{B}_{1} & \boldsymbol{C}_{2}^{\top} \boldsymbol{B}_{2} & & \\
            \vdots & \vdots & \ddots \\
            \boldsymbol{C}_{i}^{\top} \boldsymbol{A}_{1:i}^{\times} \boldsymbol{B}_{0} & \boldsymbol{C}_{i}^{\top} \boldsymbol{A}_{2:i}^{\times} \boldsymbol{B}_{1} & \cdots & \boldsymbol{C}_{i}^{\top} \boldsymbol{B}_{j}
        \end{bmatrix},
        \end{aligned}
    \end{equation}
    In this discussion, we assume the condition where matrices $\boldsymbol{A}$ and $\boldsymbol{B}$ are not subjected to discretization, 
    allowing us to assert that matrix $\boldsymbol{M}_{ij} = \boldsymbol{C}_{i}^{\top} \boldsymbol{A}_{i - j:i}^{\times} \boldsymbol{B}_{j}$. 
    The aforementioned equation delineates the transformation process of the input sequence $\boldsymbol{X} \in \mathbb{R}^{T \times D}$ into the output sequence $\boldsymbol{Y} \in \mathbb{R}^{T \times D}$ facilitated by the parameter matrices $\boldsymbol{A} \in \mathbb{R}^{T \times N}$, $\boldsymbol{C} \in \mathbb{R}^{T \times N}$, $\boldsymbol{B} \in \mathbb{R}^{T \times N}$ and $\boldsymbol{M} \in \mathbb{R}^{T \times T}$.
    The inefficiency of SSM in executing matrix operations stems from the computational demands constrained by the recursive computational load of matrix $\boldsymbol{A}_{m:n}^{\times}$ in a temporal sequence.
    Consequently, by simplifying matrix $\boldsymbol{A}_{m:n}^{\times}$ to a scalar and leveraging the properties of semi-separable matrices, we can significantly enhance the efficiency of matrix computations involving matrix $\boldsymbol{M}$.
    Under the assumption that $\boldsymbol{A}=[1, a_1, \dots, a_{t - 1}] \in \mathbb{R}^{T}$, SSD facilitates the computation of matrix $\boldsymbol{M}$ by introducing a 1-Semi Separable (1-SS) matrix $\boldsymbol{L} \in \mathbb{R}^{T \times T}$ defined as follows:
    \begin{equation}
        \label{eq:ssd_mask}
        \boldsymbol{L} = \begin{bmatrix}
            1 \\
            a_1 & 1 \\
            a_2 a_1 & a_2 & 1 \\
            \vdots & \vdots & \ddots & \ddots \\
            a_{t - 1} \ldots  a_1 & a_{t - 1} \ldots a_2 & \cdots & a_{t - 1} & 1
        \end{bmatrix}, 
    \end{equation}
    \begin{equation*}
        \boldsymbol{M}_{i + 1, j + 1} = \boldsymbol{C}_{i}^{\top} \boldsymbol{A}_{i:j} \boldsymbol{B}_j \coloneqq \boldsymbol{C}_{i}^{\top} a_i \ldots a_{j + 1} \boldsymbol{B}_j, 
    \end{equation*}
    Thus, we can equivalently derive the following equation:
    \begin{equation}
        \boldsymbol{M} = \boldsymbol{L} \circ \boldsymbol{C} \boldsymbol{B}^{\top} , 
    \end{equation}
    \begin{equation*}
        \boldsymbol{M} = 
        \begin{bmatrix}
            \boldsymbol{C}_{0}^{\top} \boldsymbol{A}_{0:0} \boldsymbol{B}_{0} & & & & \\
            \boldsymbol{C}_{1}^{\top} \boldsymbol{A}_{1:1} \boldsymbol{B}_{0} & \boldsymbol{C}_{1}^{\top} \boldsymbol{A}_{1:1} \boldsymbol{B}_{1} & & & \\
            \boldsymbol{C}_{2}^{\top} \boldsymbol{A}_{2:1} \boldsymbol{B}_{0} & \boldsymbol{C}_{2}^{\top} \boldsymbol{A}_{2:1} \boldsymbol{B}_{1} & \boldsymbol{C}_{2}^{\top} \boldsymbol{A}_{2:2} \boldsymbol{B}_{2} & & \\
            \vdots & \vdots & \ddots \\
            \boldsymbol{C}_{i}^{\top} \boldsymbol{A}_{i:1} \boldsymbol{B}_{0} & \boldsymbol{C}_{i}^{\top} \boldsymbol{A}_{i:2} \boldsymbol{B}_{1} & \cdots & \boldsymbol{C}_{i}^{\top} \boldsymbol{A}_{i:i} \boldsymbol{B}_{i}
        \end{bmatrix} . 
    \end{equation*}
    By considering matrix $\boldsymbol{C}$ as matrix $\boldsymbol{Q}$, matrix $\boldsymbol{B}$ as matrix $\boldsymbol{K}$ and matrix $\boldsymbol{X}$ as matrix $\boldsymbol{V}$ within the framework of the attention mechanism and excluding the Softmax computation, a profound conclusion is reached:
    \begin{equation}
        \boldsymbol{Y} = \boldsymbol{L} \circ \boldsymbol{C} \boldsymbol{B}^{\top} \boldsymbol{X} = \boldsymbol{L} \circ \boldsymbol{Q} \boldsymbol{K}^{\top} \boldsymbol{V} .
    \end{equation}
    This concept is referred to as Duality in the SSD framework \cite{SSD2024}. 
    By referring to the tensor contraction order in the calculation of linear attention \cite{LinearAttention2020}—specifically using the associative property of matrix multiplication where $K^T V$ is computed first 
    and subsequently left-multiplied by $Q$—the computational complexity is reduced from quadratic to linear in the sequence length without considering causal masks, a property that SSD also achieves.
    \begin{equation*}
        \boldsymbol{Y} = \left(\boldsymbol{Q} \boldsymbol{K}^{T}\right) \boldsymbol{V} \quad \Rightarrow \quad \boldsymbol{Y} = \boldsymbol{Q} \left(\boldsymbol{K}^{T} \boldsymbol{V}\right), 
    \end{equation*}
    \begin{equation*}
        \boldsymbol{Y} = \left(\boldsymbol{C} \boldsymbol{B}^{T}\right) \boldsymbol{X} \quad \Rightarrow \quad \boldsymbol{Y} = \boldsymbol{C} \left(\boldsymbol{B}^{T} \boldsymbol{X}\right), 
    \end{equation*}
    To integrate causal masked matrix without compromising linear computational complexity, 
    the block properties of the 1-SS matrix $\boldsymbol{L}$ can be leveraged in conjunction with segment accumulation. 
    This approach facilitates the implementation of causal linear attention, 
    referred to as Structured Masked Attention (SMA) \cite{SSD2024} in SSD.
    Since the SSD architecture is derived from the Mamba model, models based on the SSD architecture are referred to as Mamba2. 
    Although SSD offers significant advantages in terms of computational load—particularly in tasks with high feature dimensions—the performance of SSD in sequential recommendation tasks is inferior compared to Mamba1. We provide a detailed analysis of this issue in the following sections.

\section{METHOD}
In this section, we offer a detailed introduction to our proposed framework, TiM4Rec. Following an overview of the overall structure of TiM4Rec, we delve into the reasons behind the performance degradation of SSD compared to SSM in low-dimensional scenarios within sequential recommendation tasks. Subsequently, we explain how our time-aware enhancement method, utilizing our proposed Time-aware Masked Matrix with linear computational complexity, compensates for this performance degradation.
    \subsection{Problem Formulation and Method Overview}
    \label{sec:overview}
        \begin{figure*}[t]
            \centering
            \includegraphics[width=0.95\textwidth]{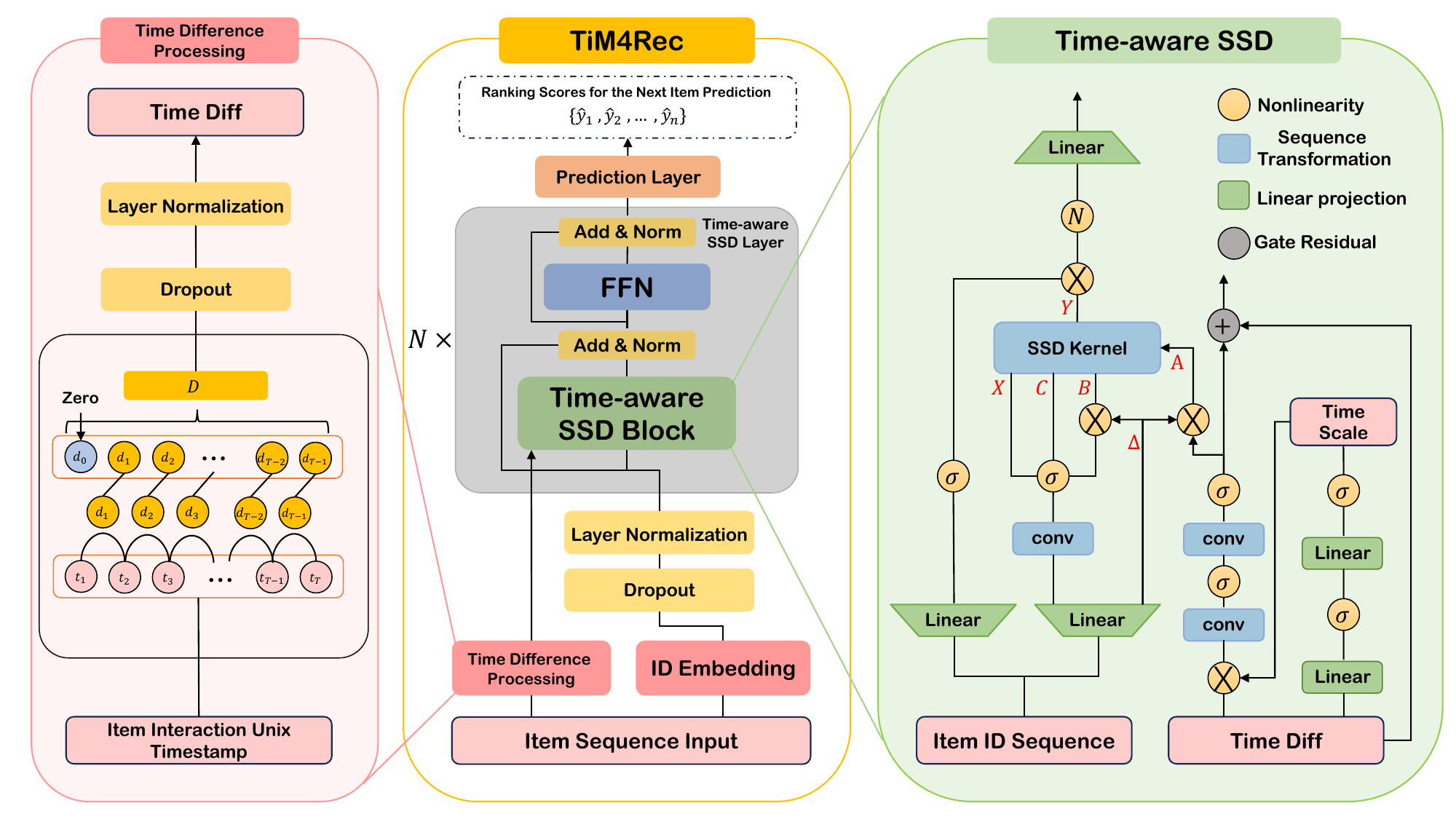} 
            \caption[The overview of TiM4Rec.]{The overview of TiM4Rec. The overall framework of TiM4Rec, which is based on time-aware enhancement, 
            enhances the performance of the SSD architecture in the domain of sequential recommendation by effectively processing the time differences in interactions between items.}
            \label{fig:TiM4RecFramework}
        \end{figure*}
        % In particular, tasks that focus solely on predicting the next item in a sequence are referred to as seq2item tasks. 
        % This work is also based on the seq2item task for its research. 
        \par
        Let $\mathcal{U} = \left\{u_1, u_2, \cdots, u_{| \mathcal{U} |}\right\}$ be the set of users and $\mathcal{V} = \left\{v_1, v_2, \cdots, v_{|\mathcal{V}|}\right\}$ be the set of items. 
        For any user $u_k$, there exists an interaction sequence $\mathcal{S}^{u_k} = \left\{v_1^{u_k}, v_2^{u_k}, \cdots, v_{T}^{u_k}\right\} \in \mathbb{R}^{T}$ order by the user interaction unix timestamp sequence $\mathcal{T}^{u_k} = \left\{t_1, t_2, \cdots, t_{T}\right\} \in \mathbb{R}^{T}$, where $v_{i}^{u_k}$ represents the i-th item of user $u_k$ interaction and $T$ representing interaction sequence length. 
        Sequential recommendation methodologies leverage the user's historical interaction sequence $\mathcal{S}^{u_k}$ to forecast the next potential interaction item $v_{T + 1}$ for the user $u_k$. 
        Without loss of generality, we extract the preference representation $p^{u_k}$ for user $u_k$ from the historical interaction sequence $\mathcal{S}^{u_k}$ and the timestamp sequence $\mathcal{T}^{u_k}$, thereby predicting the probability distribution $\hat{\mathcal{Y}}\left(v_{T+1} | p^{u_k}\right)$ for the next potential interaction item. For convenience in the subsequent discussion, the superscript $u_k$ be omitted.
        \par
        As illustrated in Fig. \ref{fig:TiM4RecFramework}, the proposed TiM4Rec is a time-aware sequential recommendation model that leverages the SSD architecture through the Time-aware SSD blocks. The TiM4Rec model consists of multiple layers of Time-aware SSD Layers, each composed of a Time-aware SSD Block and a Feed Forward Network (FFN) \cite{Transformer2017}. We incorporate a special time difference processing method that allows the SSD kernel to integrate time difference information.
         
    \subsection{Data Preprocessing}
    \subsubsection{Embedding Layer}
    In a manner analogous to previous models, we establish a learnable embedding table $\mathbb{E} = \left\{E_1, E_2, \cdots, E_{|\mathcal{V}|}\right\} \in \mathbb{R}^{|\mathcal{V}| \times D}$, the item IDs are converted from a sparse vector representation to a dense vector representation, where $E_j$ denotes the potential dense vector representation of the item $v_j$. The interaction sequence $\mathcal{S}$ can be transformed into an input sequence $\boldsymbol{X} \in \mathbb{R}^{T \times D}$ composed of dense vectors by the embedding table $\mathbb{E}$. 
    Subsequently, we enhance the embedding table $\mathbb{E}$ representation learning process by employing dropout \cite{Dropout2014} and layer normalization \cite{LayerNorm2016} techniques. These methods mitigate overfitting and stabilize the training dynamics, thereby improving the robustness and generalization capabilities of the learned embeddings.
    \subsubsection{Time Difference Processing}
        To integrate the interactive timestamp sequence $\mathcal{T}$ into the SSD computation kernel and enhance the model's ability to capture latent user interests, we perform differential processing on $\mathcal{T}$. The specific mechanisms of this integration will be discussed shortly. The interactive timestamp sequence $\mathcal{T}$ undergoes a displacement subtraction, with an initial zero appended, resulting in the time difference sequence $\mathcal{D} = \left\{d_0, d_1, \cdots, d_{T-1}\right\}\in \mathbb{R}^{T}$. The computation process is as follows: 
        \begin{equation}
            \label{eq:time_diff}
            \mathcal{D}_{i} = \begin{cases}
                0, \quad &i = 0 \\
                \mathcal{T}_{i + 1} - \mathcal{T}_{i}, \quad &i \in \left[1, T - 1\right]
            \end{cases}. 
        \end{equation}
        Subsequently, we apply dropout \cite{Dropout2014} and layer normalization \cite{LayerNorm2016} to sequence $\mathcal{D}$ in order to eliminate data noise and optimize the data distribution for subsequent analysis.

    \subsection{Time-Aware Linear Complexity Enhancement}
    \subsubsection{Analyzing the Masked Matrix in SSD}
    \label{sec:mask-ssd}
\begin{figure}[ht]
            \centering
            \includegraphics[width=0.5\textwidth]{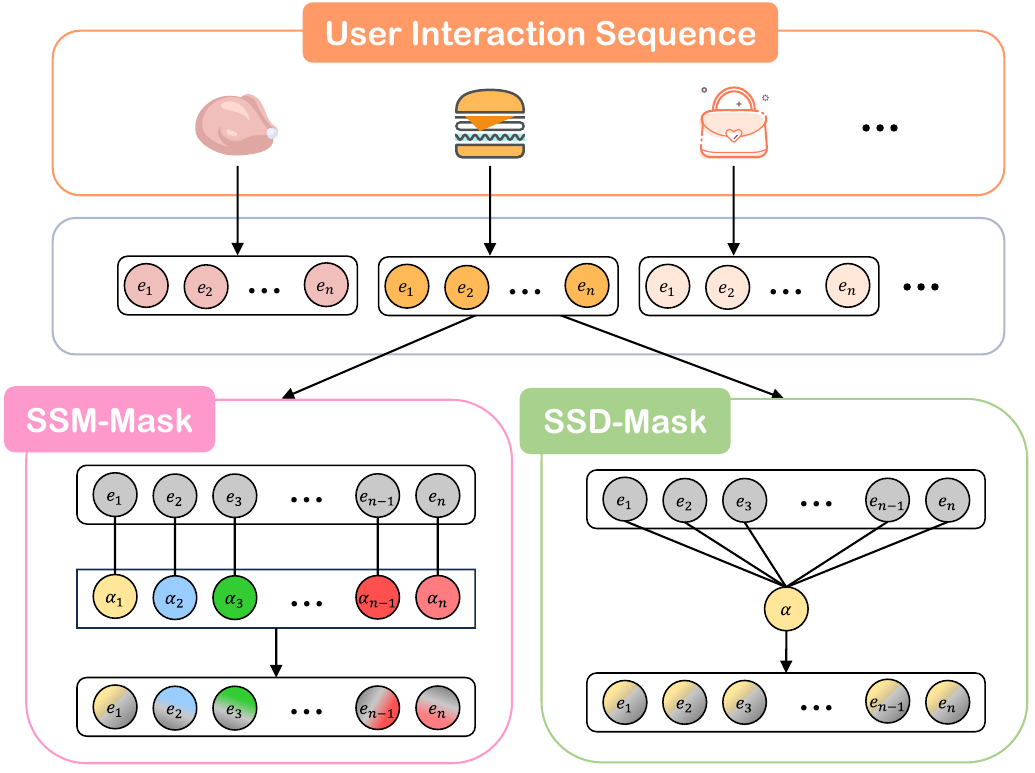} 
            \caption[mask]{The distinction in mask formulation between SSD and SSM is noteworthy. For an individual item, SSM can be viewed as an equivalent point-to-point mask, whereas SSD resembles the global mask utilized in Transformer architectures.}
            \label{fig:mask}
        \end{figure}

        As discussed in section \ref{sec:ssd}, the critical advancement from SSM to SSD resides in the scalarization of the transition matrix $\boldsymbol{A}$. 
        When considering this modification from an attention perspective, the central aspect involves the alterations in the masked matrix $\boldsymbol{L}$. 
        Although the recursive operational constraints of matrix $\boldsymbol{A}$ in SSM prevent the direct and precise formulation of the masked matrix within Eq. \ref{eq:ssm_m}, 
        it remains feasible to examine an equivalent expression for the final masked matrix:  
        \begin{equation}
            \boldsymbol{L}_{\text{SSM}} = \begin{bmatrix}
                I \\
                \boldsymbol{A}_1 & I \\
                \boldsymbol{A}_2 \boldsymbol{A}_1 & \boldsymbol{A}_2 & I \\
                \vdots & \vdots & \ddots & \ddots \\
                \boldsymbol{A}_{1:i}^{\times} & \boldsymbol{A}_{2:i}^{\times} & \cdots & \boldsymbol{A}_{i - 1:i}^{\times} & I
            \end{bmatrix}, 
        \end{equation}
        As illustrated in Fig. \ref{fig:mask}, in contrast to the masked matrix of SSD presented in Eq. \ref{eq:ssd_mask}, 
        the fundamental change in the masked matrix $L_{\text{SSM}}$ is that its mask coefficients are applied in a point-to-point manner on the individual item $\boldsymbol{E}_{i} = \left[e_1, e_2, \cdots, e_{n}\right] \in \mathbb{R}^{D}$ (The $\boldsymbol{A}$ matrix follows its definitions in SSM and SSD, respectively.):
        \begin{equation*}
            \begin{aligned}
                &\text{SSM:}\quad \hat{\boldsymbol{E}}_i = \boldsymbol{A}^{\times}_{p:q}  \cdot \boldsymbol{E}_{i}, \quad \boldsymbol{A}^{\times}_{p:q} \in \mathbb{R}^{D}, \\
                &\text{SSD:}\quad \hat{\boldsymbol{E}}_i = \boldsymbol{A}_{p:q} \cdot \boldsymbol{E}_{i}, \quad \boldsymbol{A}_{p:q} \in \mathbb{R}^{1}, 
            \end{aligned}
        \end{equation*}
        In instances where the $\boldsymbol{E}$ is in low dimension, its relative contribution to information is diminished, necessitating a more refined filtering of $\boldsymbol{E}$ by the masked matrix. 
        Consequently, the point-to-point masking employed in SSM demonstrates superior performance in low-dimensional contexts. 
        Notably, from the perspective of masking strategies, SSD exhibits a greater inclination towards the Transformer architecture than SSM. 
        \begin{gather*}
            \boldsymbol{L}_{MSA} = \begin{bmatrix}
                1 \\
                1 & 1 \\
                1 & 1 & 1 \\
                \vdots & \vdots & \ddots & \ddots \\
                1 & 1 & \cdots & 1 & 1
            \end{bmatrix}, \\
            \text{\small MSA stands for Vanilla Masked Self Attention Mechanism}
        \end{gather*}
        However, owing to the 1-SS matrix structure inherent in the masked matrix of SSD, there is a pronounced attenuation effect on the interaction items that appear earlier in the sequence. 
        This feature aligns SSD more closely with the actual user interaction patterns observed in real-world recommendation systems. 
        Consequently, it is anticipated that, under equivalent conditions in low-dimensional settings, the SSM architecture exhibit superior performance in the realm of sequential recommendation, followed by SSD, with the Transformer architecture trailing behind.
        This hypothesis is further corroborated by the findings presented in our subsequent experimental section. 
        \par
        Observing the differences between the equivalent masked matrices of SSD and SSM, 
        it can be seen that the core of their mask matrices is recursively generated based on the state transition matrix $\boldsymbol{A}$. As described in section \ref{sec:ssd}, 
        the only difference is that SSD achieves more efficient matrix multiplication calculations by scalar quantization of matrix $\boldsymbol{A}$. 
        Therefore, addressing the loss of the masked matrix caused by the scalar quantization of matrix $\boldsymbol{A}$ while maintaining the efficient matrix multiplication calculation of SSD has become the key to improvement.
        
        \subsubsection{Time-aware Structured Masked Matrix}
        \label{sec:TSMM}
        Let's reconsider the time-aware enhancement approach of TiSASRec \cite{TiSASRec2020}. The essence of this method lies in embedding the time difference matrix and processing it through the attention mechanism to generate a time-aware attention score matrix, which is then integrated into the final attention score matrix. \textbf{If we regard the generated time-aware attention score matrix as a special type of masked matrix, it aligns with our previous analysis: employing a more refined mask matrix to extract potential user interest features can lead to improved recommendation performance.}
        However, in the TiSASRec method, the computational steps involved in generating the time-aware masked matrix by embedding the time difference matrix and processing it through the attention mechanism incur substantial computational overhead. Therefore, it is imperative to identify a direct and efficient approach to convert the time difference matrix into a time-aware masked matrix.
        \par
        For a sequence of interactive timestamps $\mathcal{T}$ (The definition is the same as in section \ref{sec:overview}), we can derive the relative time difference matrix $\hat{\mathcal{D}}$ between each product through the following transformation: 
        \begin{equation}
            \label{eq:time_diff_matrix}
            \hat{\mathcal{D}} = \begin{bmatrix}
                \hat{d}_{11} \\
                \mathbf{\underline{\hat{d}_{21}}} & \hat{d}_{22}\\
                \hat{d}_{31} & \mathbf{\underline{\hat{d}_{32}}} & \hat{d}_{33}\\
                \vdots & \vdots & \ddots & \ddots \\
                \hat{d}_{T1} & \hat{d}_{T2} & \cdots & \mathbf{\underline{\hat{d}_{T T-1}}} & \hat{d}_{TT}
            \end{bmatrix}, \quad \hat{d}_{mn} = t_m - t_n, 
        \end{equation}
        Directly applying matrix $\hat{\mathcal{D}}$ to the masked matrix $\boldsymbol{L}$ (as described in Eq. \ref{eq:ssd_mask}) compromises the 1-SS matrix property of $\boldsymbol{L}$. 
        This compromise prevents us from leveraging the efficient matrix computation techniques offered by the SSD architecture, 
        consequently resulting in a return to quadratic computational complexity. Therefore, it is imperative to modify the insertion method of matrix $\hat{\mathcal{D}}$.
        \par
        When we construct a new matrix $\mathcal{D}$ by selecting only the values of the underlined elements in matrix $\hat{\mathcal{D}}$, we effectively adopt the form described in Eq. \ref{eq:time_diff}. This approach allows us to integrate matrix $\mathcal{D}$ into the masked matrix $\boldsymbol{L}$ with linear complexity. 
        Initially, we apply matrix $\mathcal{D}$ to the scalarization matrix $\boldsymbol{A}$, which results in the corresponding positions in the masked matrix $\boldsymbol{L}$ containing the relevant time difference information 
        (Due to the presence of layer normalization, the actual value of $d_1$ will be non-zero.):
        \begin{equation*}
            \hat{\boldsymbol{A}}_i = \hat{a}_i = \mathcal{D}_i \cdot \boldsymbol{A}_i = \hat{d}_{i i-1} \cdot a_i = d_{i} \cdot a_{i}, 
        \end{equation*}
        \begin{equation*}
            \hat{\boldsymbol{A}}_{m:n} \coloneqq \hat{a}_n \hat{a}_{n - 1} \cdots \hat{a}_{m} = a_n d_n \cdot a_{n-1} d_{n-1} \cdots a_{m} d_{m},
        \end{equation*}
        \begin{equation}
            \label{eq:mask_time}
            \boldsymbol{L} = \begin{bmatrix}
                \hat{a}_0 \\
                \hat{a}_1 & \hat{a}_0 \\
                \hat{a}_2 \hat{a}_1 & \hat{a}_2 & \hat{a}_0 \\
                \vdots & \vdots & \ddots & \ddots \\
                \hat{a}_{t - 1} \ldots  \hat{a}_1 & \hat{a}_{t - 1} \ldots \hat{a}_2 & \cdots & \hat{a}_{t - 1} & \hat{a}_0
            \end{bmatrix}, 
        \end{equation}
        We refer to $\boldsymbol{L}$ as a \textbf{Time-aware Structured Masked Matrix}. This can be regarded as an equivalent approximation of $\hat{d}_{ij}$ as follows (Notably, the computation in Eq. \ref{eq:mask_time} employs the segment accumulation, consistent with the 1-SS matrix computation in SSD \cite{SSD2024}, to maintain efficient computational performance.):
        \begin{equation*}
            \hat{d}_{ij} \approx \hat{d}_{i i-1} \hat{d}_{i-1 i-2} \cdots \hat{d}_{j + 1 j} = d_i d_{i-1} \cdots d_{j}, 
        \end{equation*}
        \par
        Through the approximation method mentioned above, we accomplish the time-aware enhancement of the SSD architecture. This is achieved without disrupting the 1-SS property of the masked matrix $\boldsymbol{L}$.
        \par
        \begin{figure}[ht]
            \centering
            \includegraphics[width=0.5\textwidth]{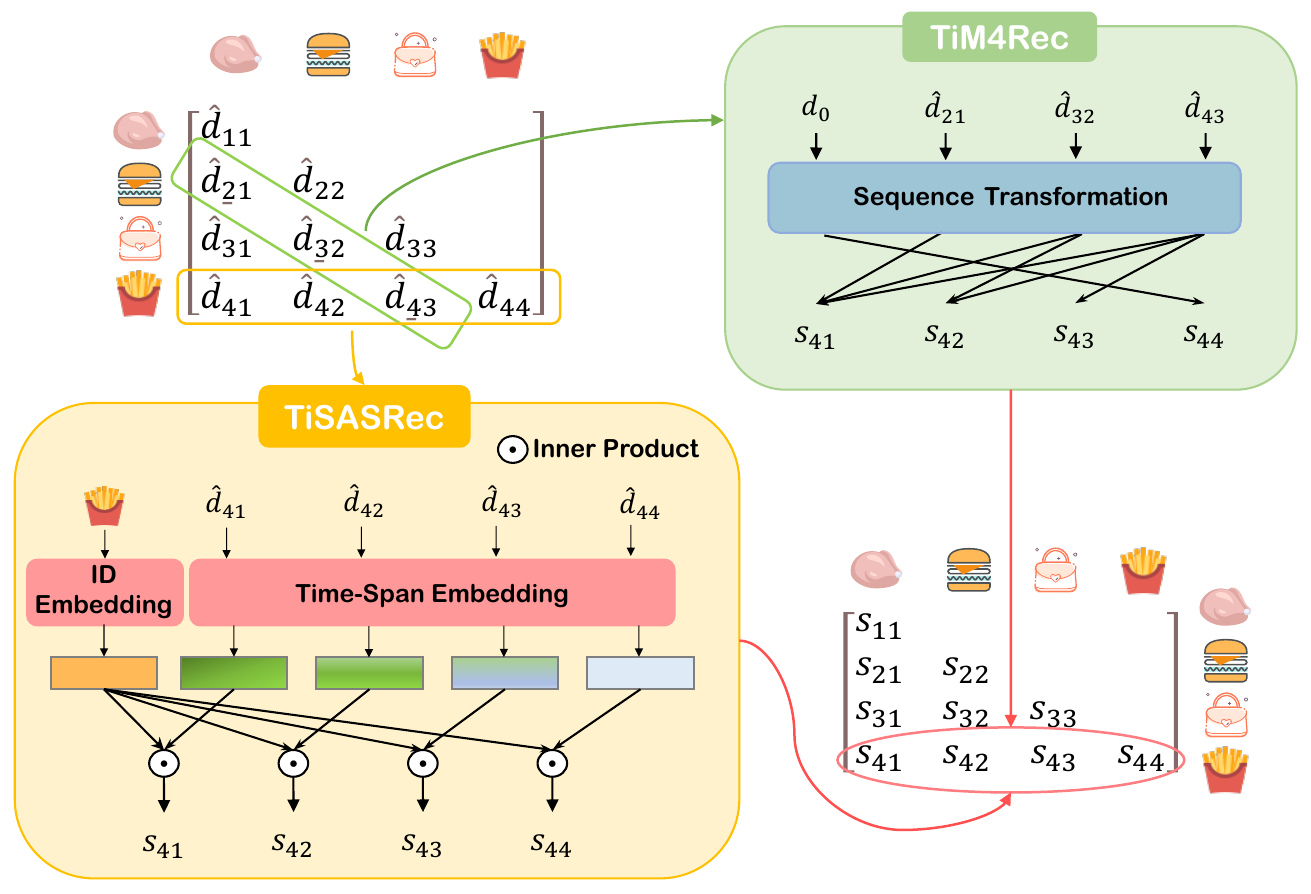} 
            \caption[time-aware]{In comparing our proposed method with the time-aware enhancement approach based on the TiSASRec paradigm, we emphasize the distinctions in how time difference information is transformed into attention score matrices. The time difference matrix is defined in the upper left corner, as outlined in Eq. \ref{eq:time_diff_matrix}.}
            \label{fig:time-aware}
        \end{figure}
        By directly integrating the time difference matrix into the masked matrix using the aforementioned method, computational efficiency is maintained. As illustrated in Figure. \ref{fig:time-aware}, unlike the time-aware augmentation methods based on the TiSASRec paradigm \cite{TiSASRec2020}, our approach circumvents the need to elevate time difference information to high-dimensional processing. Instead, it employs a straightforward scalar-level sequence transformation (which will be detailed shortly) to effectively capture personalized user interaction patterns. This enhanced time difference information is then utilized to construct a Time-aware Structured Masked Matrix $\boldsymbol{L}$ that adheres to the property of 1-SS matrix, integrating seamlessly into the SSD kernel. This approach addresses the dual challenges of computational efficiency and integration into the SSD kernel.
        
        \subsection{Time-aware SSD Block}
        In the following section, we will elaborate on the integration of the Time-aware Structured Masked Matrix into the SSD kernel.
        Despite providing a detailed derivation of the SSD in Section \ref{sec:ssd}, 
        the actual code implementation of the SSD kernel retains the discretization process. 
        Therefore, further analysis is required. 
        Notably, in Section \ref{sec:ssm}, we derive the discretization for the SSM, 
        but it is important to highlight that the actual computational process is approximated in both SSM and SSD implementations. 
        Specifically, for matrix $\overline{\boldsymbol{A}}$, the implementation approximates the exponential operation by directly computing $\overline{\boldsymbol{A}} \approx \Delta \boldsymbol{A}$. 
        Meanwhile, for matrix $\overline{\boldsymbol{B}}$, given the Maclaurin series expansion of $\exp(x) $ as $ 1 + x + \frac{x^2}{2} + \cdots $, the following approximation can be made: 
        \begin{equation}
            \begin{aligned}
                \overline{\boldsymbol{B}} &\approx \left(\Delta \boldsymbol{A}\right)^{-1} \left(\boldsymbol{I} + \Delta \boldsymbol{A} - \boldsymbol{I}\right) \cdot \Delta \boldsymbol{B} \\
                & = \left(\Delta \boldsymbol{A}\right)^{-1} \Delta \boldsymbol{A} \cdot \Delta \boldsymbol{B} \\
                & = \Delta \boldsymbol{B}, 
            \end{aligned}
        \end{equation}
        % In this approximation, we establish the following definitions: 
        % the state transition coefficient $ \boldsymbol{A} \in \mathbb{R}^{1} $. 
        Under this approximation, the original state transition matrix $\boldsymbol{A} \in \mathbb{R}^{T}$ in the SSD can be transformed into a scalar coefficient $A \in \mathbb{R}^{1}$, which when multiplied by the discretization parameter $\Delta \in \mathbb{R}^{T}$, yields the discretized state transition matrix $\overline{\boldsymbol{A}} \in \mathbb{R}^{T}$. This approximation process optimizes computational efficiency while maintaining maximal accuracy. Given that this approach is empirically established by Gu et al. \cite{SSD2024}, we will not delve into further details here.
        \par
        Building upon the aforementioned consensus, we now proceed to a comprehensive derivation of the Time-aware SSD Block method.
        By inputting $ \boldsymbol{X} \in \mathbb{R}^{T \times D} $, we derive matrices $ \boldsymbol{B} \in \mathbb{R}^{T \times N} $ and $ \boldsymbol{C} \in \mathbb{R}^{T \times N} $, along with the discretization parameter $ \Delta \in \mathbb{R}^{T} $: 
        \begin{equation}
            \boldsymbol{X}, \boldsymbol{B}, \boldsymbol{C}, \Delta= \boldsymbol{X} W  + b, 
        \end{equation}
        Where the weight matrix $ W \in \mathbb{R}^{D \times (D + 2N + 1)} $ and the bias matrix $ b \in \mathbb{R}^{D + 2N + 1} $. 
        Subsequently, a causal convolution transformation is applied to the matrices $ \boldsymbol{X} $, $ \boldsymbol{B} $ and $ \boldsymbol{C} $:
        \begin{gather}
            \boldsymbol{X}_t, \boldsymbol{B}_t, \boldsymbol{C}_t = \sigma\left[(\boldsymbol{X}_t, \boldsymbol{B}_t, \boldsymbol{C}_t)^{\top} * \omega \right],  \\
            \text{where} \quad \mathcal{Q}_t = \mathcal{P}_t * \omega \coloneqq \sum_{m=0}^{K-1} \mathcal{P}_{\max\left(t - m, 0\right)} \cdot \omega_{m}, \notag
        \end{gather}
        Where the $\omega \in \mathbb{R}^{K}$ is convolution kernel and the $\sigma$ represents non-linear activation functions. 
        \par
        Before proceeding with the derivation, it is prudent to examine how the time difference vector $\mathcal{D}$ is integrated into the discretized SSD. 
        Regarding the time difference matrix $\mathcal{D}$, we note that the implicit interest conveyed by the same interaction time difference can vary across different user interaction contexts. 
        Therefore, we initially apply a time-varying transformation to the time difference matrix using the following transformation:
        \begin{equation}
            \begin{aligned}
                \alpha^{\mathcal{D}} &= \sigma \left[\sigma \left(\mathcal{D} W_{1}^{\mathcal{D}} + b_{1}^{\mathcal{D}}\right) W_{2}^{\mathcal{D}} + b_{2}^{\mathcal{D}}\right], \\
                \mathcal{D} &= \alpha^{\mathcal{D}} \cdot \mathcal{D}, 
            \end{aligned}
        \end{equation}
        Where the weight matrix $W_{1}^{\mathcal{D}},  W_{2}^{\mathcal{D}} \in \mathbb{R}^{T \times T}$ and the bias matrix $  b_{1}^{\mathcal{D}},  b_{2}^{\mathcal{D}} \in \mathbb{R}^{T} $. 
        Considering the convergence of interaction time differences within a short time frame, we apply causal convolution to the time difference matrix $\mathcal{D}$ to enhance the implicit features of user interest points.
        \begin{equation}
            \mathcal{D}_t = \sigma\left(\mathcal{D}_t * \omega^{\mathcal{D}}\right) \coloneqq \sigma\left(\sum_{m=0}^{K-1} \mathcal{D}_{\max\left(t - m, 0\right)} \cdot \omega^{\mathcal{D}}_{m} \right), 
        \end{equation}
        Next, we integrate the enhanced time difference matrix $\mathcal{D}$ with the discrete parameter $\Delta$, 
        and obtain matrices $ \overline{\boldsymbol{A}} $ and $ \overline{\boldsymbol{B}} $ through the following transformation: 
        \begin{equation}
            \hat{\Delta} = \Delta \cdot \mathcal{D}, \quad
            \overline{\boldsymbol{A}} = A \cdot \hat{\Delta}, \quad
            \overline{\boldsymbol{B}} = \hat{\Delta} \cdot \boldsymbol{B}, 
        \end{equation}
        Following the methodology outlined in Section \ref{sec:TSMM}, we employ the following transformations to construct the Time-aware Structured Masked Matrix $\boldsymbol{L}$:
        \begin{equation*}
            \hat{\boldsymbol{A}}_i = \hat{a}_i = \overline{\boldsymbol{A}}_i = A \cdot \hat{\Delta}_{i} = A \cdot \Delta_i \cdot d_i, 
        \end{equation*}

        \begin{equation}
            \boldsymbol{L} = \begin{bmatrix}
                \hat{a}_0 \\
                \hat{a}_1 & \hat{a}_0 \\
                \hat{a}_2 \hat{a}_1 & \hat{a}_2 & \hat{a}_0 \\
                \vdots & \vdots & \ddots & \ddots \\
                \hat{a}_{t - 1} \ldots  \hat{a}_1 & \hat{a}_{t - 1} \ldots \hat{a}_2 & \cdots & \hat{a}_{t - 1} & \hat{a}_0
            \end{bmatrix}, 
        \end{equation}
        Finally, the following equation can be derived to map the input sequence $\boldsymbol{X}$ and $\mathcal{D}$ to the output $\boldsymbol{Y} \in \mathbb{R}^{T \times D} $: 
        % (TSMM $\boldsymbol{L}$ refers to Eq. \ref{eq:mask_time})
        \begin{equation}
            \boldsymbol{Y} = TiSSD\left(\boldsymbol{X}, \mathcal{D}\right) \coloneqq \boldsymbol{L} \circ \boldsymbol{C} \overline{\boldsymbol{B}}^{\top} \boldsymbol{X} = \boldsymbol{L} \circ \boldsymbol{C} \boldsymbol{B}^{\top} \left(\hat{\Delta}^{\top} \boldsymbol{X}\right), 
        \end{equation}
        Due to the preservation of the 1-SS matrix property of $\boldsymbol{L}$, the matrix computation acceleration method discussed in Section \ref{sec:ssd} remains applicable.
        \par
        Multiple layers of Time-aware SSD blocks are stacked to extract deeper user interest features. Consequently, after the Time-aware SSD, the features are transformed by adding an FFN \cite{Transformer2017} layer to adapt to the next semantic space.
        It is noteworthy that to adapt the time difference vector $\mathcal{D}$ to the feature semantic space of the next layer, gate residual \cite{ResNet2016} processing is applied to the input time difference vector $\mathcal{D}$ for the subsequent layer.
        
        \subsection{Prediction Layer}
        By utilizing the Time-aware SSD Layer, input $\boldsymbol{X}$ can be transformed into a user interest feature sequence $\boldsymbol{P} \in \mathbb{R}^{T \times D}$. 
        The last element $p \in \mathbb{R}^{D}$ of the feature sequence $\boldsymbol{P}$ is then extracted as the interest feature corresponding to the current user interaction sequence.
        Afterward, we compute the dot product between the user interest feature $p$ and the embeddings of all items $\mathbb{E} \in \mathbb{R}^{|\mathcal{V}| \times D}$. This result is then scaled using the Softmax function to obtain the interaction probability scores $\hat{\mathcal{Y}}\left(v_{T+1} | p\right) \in \mathbb{R}^{|\mathcal{V}|}$ for each item from the user.
        \begin{equation}
            \hat{\mathcal{Y}}\left(v_{T+1} | p\right) = Softmax\left(p \cdot \mathbb{E}^T\right), 
        \end{equation}
        % As stated in Section 2.1, the predicted scores $\hat{\mathcal{Y}} \in \mathbb{R}^{|\mathcal{V}|}$ for all items are obtained by calculating the inner product of $p_t$ and the embedding table of all items. However, a notable difference is that the actual score calculation incorporates the Softmax function, 
        % even though the Softmax operation does not influence the ranking of the scores.
        In alignment with previous work \cite{SASRec2018}, we optimize the model parameters using the Cross Entropy loss function. Let us denote the true interaction distribution for user $u_k$ as $g^{u_k} \in \mathbb{R}^{|\mathcal{V}|}$; then, the following loss function is defined:
        \begin{equation}
            \mathcal{L} = - \sum_{u_k}^{|\mathcal{U}|} \sum_{i}^{|\mathcal{V}|} g^{u_k}_{i} \ln\left(\hat{\mathcal{Y}}^{u_k}_{ i}\right)
        \end{equation}
        
        \subsection{Complexity Analysis}
        \label{sec:complexity}
        Models based on the Vanilla Transformer \cite{Transformer2017}, such as SASRec \cite{SASRec2018}, require a total of $O(L^2 N)$ floating-point operations (FLOPs). 
        Models based on Linear Recurrent Units \cite{LRU2023}, such as LRURec \cite{LRURec2024}, require $O(\log(T)N^2)$ FLOPs.
        In contrast, thanks to the computational improvements provided by Mamba, models based on SSM \cite{Mamba2023} and SSD \cite{SSD2024}, such as Mamba4Rec \cite{Mamba4Rec2024} and SSD4Rec \cite{SSD4Rec2024}, only require $O(T N^2)$ FLOPs. Notably, although the computational complexity of the SSM and SSD architectures is identical, the SSD architecture can efficiently leverage matrix multiplication for computations, resulting in faster actual training times compared to the SSM architecture.
        In practical application scenarios of sequential recommendation, including news and product recommendations, it is often the case that $T \gg  N$. Thus, sequential recommendation models based on Mamba demonstrate superior advantages. 
        % TiSASRec incorporates time-awareness through time difference encoding; however, due to the inherent design drawbacks of its time difference encoding matrix calculation, TiSASRec necessitates a total of $O(L^2 N^2)$ FLOPs. 
        TiSASRec \cite{TiSASRec2020} enhances its performance by integrating time-aware capabilities via time difference encoding. Despite the theoretical FLOPS remaining in the order of $O(T^2 N)$, the practical computational and memory demands exceed those of SASRec by more than four times. This substantial increase is attributed to the inherent computational constraints associated with time difference encoding. Consequently, while TiSASRec demonstrates improved time sensitivity, it necessitates a significantly higher computational overhead and memory allocation compared to SASRec, posing potential challenges in scalability and efficiency.
        % In contrast, TiM4Rec retains the benefits of the SSD architecture by employing a time difference masked matrix with linear computational complexity, still requiring only $O(L N^2)$ FLOPs.
        In contrast, TiM4Rec preserves the benefits of the SSD architecture by employing a Time-aware Structured Masked Matrix characterized by linear computational complexity. This method preserves computational efficiency, requiring only $O(TN^2)$FLOPs, and demands only an additional $O(T)$ level of FLOPs.
        \par
        Notably, the generation of the time-aware masked matrix in TiM4Rec is solely based on scalar computations. This characteristic ensures that the additional computational complexity does not escalate in high-dimensional modeling scenarios, thereby preserving the superior matrix operation efficiency inherent to the SSD4Rec architecture. 
        Our summarized results are shown in Table \ref{tab:complexity}. 
        \begin{table}[ht]
            \centering
            \caption{Comparison of computational complexity between representative sequential recommendation methods, where "M.M." refers to "Matrix Multiplication", "T.F" represents "Time-aware additional FLOPs"}
            \setlength{\tabcolsep}{0.5em}{
            \resizebox{0.45\textwidth}{!}{
              \begin{tabular}{ccccc}
              \toprule
                    & Architecture & FLOPs & T.F & M.M.\\
              \midrule
              SASRec & Attention & $T^2N$ & - & \checkmark \\
              LRURec & LRU & $\log(T) N^2$ & - &  \\
              Mamba4Rec & SSM & $TN^2$ & - &  \\
              SSD4Rec & SSD & $TN^2$ & - & \checkmark \\
              TiSASRec & Attention & $T^2 N$ & $T^2 N$ & \checkmark \\
              TiM4Rec & SSD & $TN^2$ & $T$ & \checkmark \\
              \bottomrule
              \end{tabular}%
              }
              }
            \label{tab:complexity}%
        \end{table}%

\section{Experiments}
  \subsection{Experimental Setup}
    \subsubsection{\textbf{Datasets}}
    The performance of our proposed model is evaluated through experiments conducted on three publicly available datasets, which have previously been utilized as evaluation benchmarks in several classical sequential recommendation models. 
    \begin{itemize}[leftmargin=10pt]
      \item \textbf{MovieLens-1M} \cite{MovieLes2016}: A dataset containing approximately 1 million user ratings for movies, collected from the MovieLens platform.
      \item \textbf{Amazon-Beauty} \cite{Amazon2014}: The user review dataset collected in the beauty category on the Amazon platform was compiled up to the year 2014.
      \item \textbf{Kuai Rand} \cite{KuaiRand2022}: Acquired from the recommendation logs of the application Kuaishou, the dataset includes millions of interactions involving items that were randomly displayed.
    \end{itemize}
    For each user, we sort their interaction records according to the timestamps, thereby generating an interaction sequence for each individual. And then retain only those users and items that are associated with a minimum of five interaction records, in accordance with the methodology established in prior research \cite{SASRec2018}. 
    The statistics of these datasets are shown in Table \ref{tab:dataset}. 
    \begin{table}[ht]
      \centering
      \caption{Statistics of the experimented datasets.}
      \vskip -0.1in
        \begin{tabular}{crrr}
        \toprule
        \textbf{Dataset} & \multicolumn{1}{c}{\textbf{ML-1M}} & \multicolumn{1}{c}{\textbf{Beauty}}  & \multicolumn{1}{c}{\textbf{KuaiRand}} \\
        \midrule
        \#Users & 6,040  & 22,363 & 23,951 \\
        \#Items & 3,416  & 12,101 & 7,111 \\
        \#Interactions & 999,611 & 198,502 & 1,134,420 \\
        Avg.Length & 165.5 & 8.9 & 47.4 \\
        Max.Length & 2,314  & 389 & 809 \\
        Sparsity & 95.15\% & 99.93\% & 99.33\% \\
        \bottomrule
        \end{tabular}
      \label{tab:dataset}
        \vskip -0.1in
    \end{table}
  \subsubsection{\textbf{Baseline}}
  To demonstrate the superiority of our proposed method, 
  we select a set of representative sequential recommendation models to serve as baselines. 
  Notably, even though SSD4Rec \cite{SSD4Rec2024} has not been released as open-source, we implemented a variant termed SSD4Rec*, adhering to the traditional sequential recommendation framework to validate the efficacy of our time-aware approach. 
  This implementation is strictly based on the SSD architecture and does not incorporate features such as variable-length sequences and bidirectional SSD methods proposed in SSD4Rec. 
  The exclusion of these features is intentional, as they are generally regarded as universal enhancement techniques, particularly in the context of handling variable-length sequences.
  \begin{itemize}[leftmargin=10pt]
    \item \textbf{Caser} \cite{TangW2018}: A classical sequential recommendation model based on CNN. 
    \item \textbf{GRU4Rec} \cite{GRU4Rec2016}: A RNN-based method constructed by Gated Recurrent Units (GRU). 
    \item \textbf{SASRec} \cite{SASRec2018}: The first exploration model of applying Transformer in the field of sequential recommendation. 
    \item \textbf{BERT4Rec} \cite{BERT4Rec2019}: A bidirectional attention sequential recommendation model following the BERT \cite{BERT2019} model paradigm. 
    \item \textbf{TiSASRec} \cite{TiSASRec2020}: A time-aware enhanced sequential recommendation model based on SASRec: Considered as a paradigm in time-aware sequential recommendation research. 
    \item \textbf{LRURec} \cite{LRURec2024}: A sequential recommendation model based on Linear Recurrent Units (LRU) \cite{LRU2023} with Parallel Scan \cite{Parallelizing2018} optimization. 
    \item \textbf{Mamba4Rec} \cite{Mamba4Rec2024}: A pioneering model that explores the application of the Mamba architecture in the domain of sequential recommendation. 
    \item \textbf{SSD4Rec} \cite{SSD4Rec2024}: The pioneering sequential recommendation model leveraging the SSD architecture, exploits its inherent advantages over the SSM architecture for high-dimensional modeling, 
    and integrates variable-length sequence training techniques to enhance model performance.
  \end{itemize}
  
  \begin{table*}[ht]
    \centering
    \caption{Sequential recommendation performance on the ML-1M, Beauty, and KuaiRand datasets. 
    Here, R@k refers to Recall@K, N@K denotes NDCG@K, and M@K represents MRR@K. 
    The best and second-best results are indicated in bold and underlined, respectively. 
    "\emph{Improv. SSD}" signifies the performance improvement of TiM4Rec compared to the SSD4Rec* results.
    "\emph{Improv. Sec}" signifies the performance improvement of TiM4Rec compared to the second-best results.
    The asterisk in SSD4Rec* denotes that this is our own replicated version. The model enclosed in boxes represents the use of time aware enhancement methods. }
    \setlength{\tabcolsep}{0.3em}{
     \resizebox{1.0\textwidth}{!}{
    \begin{tabular}{cclcccccccccrr}
    \toprule
    \multicolumn{2}{l}{\makecell{Model Type \& Model $\rightarrow$}} &  & CNN & RNN & \multicolumn{3}{c}{Transformer} & LRU & SSM & \multicolumn{2}{c}{SSD} & \multirow{2}{*}{\makecell{\emph{Improv.} \\ SSD}} & \multirow{2}{*}{\makecell{\emph{Improv.} \\ Sec}}\\
    \cmidrule(l){1-2} \cmidrule(l){4-4} \cmidrule(l){5-5} \cmidrule(l){6-8} \cmidrule(l){9-9} \cmidrule(l){10-10} \cmidrule(l){11-12}
    Dataset & Metric &  & {\small Caser} & {\small GRU4Rec} & {\small SASRec} & {\small BERT4Rec} & \fbox{{\small TiSASRec}} & \small{LRURec} & {\small Mamba4Rec} & {\small SSD4Rec*} & \fbox{{\small TiM4Rec}} & \\
    \hline
    \midrule
    \multirow{9}{*}{\makecell{MovieLens-1M}} 
    & R@10 & & 0.2156 & 0.2985 & 0.3060 & 0.2800 & 0.3147 & 0.3002 & \underline{0.3253} & 0.3199 & \textbf{0.3310} & \cellcolor[HTML]{cce6ff}3.47\% & \cellcolor[HTML]{EFFBEC}1.75\% \\
    & R@20 & & 0.3083 & 0.3937 & 0.4050 & 0.3853 & 0.4250 & 0.4035 & \textbf{0.4354} & 0.4240 & \underline{0.4338} & \cellcolor[HTML]{cce6ff}2.31\% & \cellcolor[HTML]{FBECEC}-0.30\% \\
    & R@50 & & 0.4515 & 0.5397 & 0.5455 & 0.5377 & 0.5608 & 0.5455 & \underline{0.5707} & 0.5649 & \textbf{0.5770} & \cellcolor[HTML]{cce6ff}2.14\% & \cellcolor[HTML]{EFFBEC}1.10\% \\
    
    & N@10 & & 0.1157 & 0.1700 & 0.1754 & 0.1584 & 0.1834 & 0.1727 & \underline{0.1891} & 0.1841 & \textbf{0.1932} & \cellcolor[HTML]{cce6ff}4.94\% & \cellcolor[HTML]{EFFBEC}2.17\% \\
    & N@20 & & 0.1391 & 0.1941 & 0.2005 & 0.1852 & 0.2106 & 0.1987 & \underline{0.2169} & 0.2104 & \textbf{0.2194} & \cellcolor[HTML]{cce6ff}4.28\% & \cellcolor[HTML]{EFFBEC}1.15\% \\
    & N@50 & & 0.1675 & 0.2232 & 0.2285 & 0.2154 & 0.2376 & 0.2268 & \underline{0.2436} & 0.2384 & \textbf{0.2477} & \cellcolor[HTML]{cce6ff}3.90\% & \cellcolor[HTML]{EFFBEC}1.68\% \\
    
    & M@10 & & 0.0853 & 0.1307 & 0.1357 & 0.1214 & 0.1424 & 0.1337 & \underline{0.1474} & 0.1425 & \textbf{0.1512} & \cellcolor[HTML]{cce6ff}6.11\% & \cellcolor[HTML]{EFFBEC}2.58\% \\
    & M@20 & & 0.0917 & 0.1373 & 0.1425 & 0.1288 & 0.1498 & 0.1407 & \underline{0.1551} & 0.1498 & \textbf{0.1584} & 
    \cellcolor[HTML]{cce6ff}5.23\%& \cellcolor[HTML]{EFFBEC}2.13\% \\
    & M@50 & & 0.0963 & 0.1420 & 0.1471 & 0.1337 & 0.1542 & 0.1452 & \underline{0.1593} & 0.1542 & \textbf{0.1629} & 
    \cellcolor[HTML]{cce6ff}5.64\%& \cellcolor[HTML]{EFFBEC}2.26\% \\
    \midrule
    \multirow{9}{*}{\makecell{Amazon \\ Beauty}} 
     & R@10 & & 0.0402 & 0.0563 & \underline{0.0851} & 0.0352 & 0.0802 & 0.0840 & 0.0838 & 0.0806 & \textbf{0.0854} & \cellcolor[HTML]{cce6ff}5.96\% & \cellcolor[HTML]{EFFBEC}0.35\% \\
     & R@20 & & 0.0618 & 0.0856 & \underline{0.1194} & 0.0563 & 0.1147 & 0.1191 & 0.1185 & 0.1146 & \textbf{0.1204} & \cellcolor[HTML]{cce6ff}5.06\% & \cellcolor[HTML]{EFFBEC}0.84\% \\
     & R@50 & & 0.1090 & 0.1376 & 0.1759 & 0.0999 & 0.1734 & 0.1799 & \textbf{0.1802} & 0.1710 & \underline{0.1800} & \cellcolor[HTML]{cce6ff}5.26\% & \cellcolor[HTML]{FBECEC}-0.11\% \\
     
     & N@10 & & 0.0199 & 0.0297 & 0.0425 & 0.0172 & 0.0405 & 0.0427 & \underline{0.0435} & 0.0423 & \textbf{0.0446} & \cellcolor[HTML]{cce6ff}5.44\% & \cellcolor[HTML]{EFFBEC}2.53\% \\
     & N@20 & & 0.0253 & 0.0371 & 0.0511 & 0.0225 & 0.0491 & \underline{0.0522} & \underline{0.0522} & 0.0509 & \textbf{0.0533} & \cellcolor[HTML]{cce6ff}4.72\% & \cellcolor[HTML]{EFFBEC}2.11\% \\
     & N@50 & & 0.0346 & 0.0473 & 0.0623 & 0.0311 & 0.0605 & 0.0640 & \underline{0.0644} & 0.0620 & \textbf{0.0651} & \cellcolor[HTML]{cce6ff}5.00\% & \cellcolor[HTML]{EFFBEC}1.09\% \\
     
     & M@10 & & 0.0138 & 0.0217 & 0.0294 & 0.0117 & 0.0282 & 0.0300 & \underline{0.0311} & 0.0306 & \textbf{0.0321} & \cellcolor[HTML]{cce6ff}4.90\% & \cellcolor[HTML]{EFFBEC}3.22\% \\
     & M@20 & & 0.0153 & 0.0237 & 0.0317 & 0.0132 & 0.0306 & 0.0326 & \underline{0.0335} & 0.0329 & \textbf{0.0345} & \cellcolor[HTML]{cce6ff}4.86\% & \cellcolor[HTML]{EFFBEC}2.99\% \\
     & M@50 & & 0.0167 & 0.0253 & 0.0335 & 0.0145 & 0.0325 & 0.0345 & \underline{0.0354} & 0.0347 & \textbf{0.0363} & \cellcolor[HTML]{cce6ff}4.61\% & \cellcolor[HTML]{EFFBEC}2.54\% \\
  
     \midrule
     \multirow{9}{*}{\makecell{Kuai Rand}} 
     & R@10 & & 0.0801 & 0.1020 & 0.1055 & 0.0938 & 0.1057 & 0.1036 & \underline{0.1094} & 0.1055 & \textbf{0.1109} & \cellcolor[HTML]{cce6ff}5.12\% & \cellcolor[HTML]{EFFBEC}1.37\% \\
     & R@20 & & 0.1344 & 0.1659 & 0.1704 & 0.1537 & 0.1710 & 0.1663 & \underline{0.1768} & 0.1717 & \textbf{0.1774} & \cellcolor[HTML]{cce6ff}3.32\% & \cellcolor[HTML]{EFFBEC}0.34\% \\
     & R@50 & & 0.2561 & 0.3017 & 0.3074 & 0.2873 & 0.3060 & 0.3078 & \underline{0.3154} & 0.3088 & \textbf{0.3202} & \cellcolor[HTML]{cce6ff}3.69\% & \cellcolor[HTML]{EFFBEC}1.52\% \\
 
     & N@10 & & 0.0395 & 0.0564 & 0.0584 & 0.0510 & 0.0590 & 0.0570 & \underline{0.0608} & 0.0588 & \textbf{0.0611} & \cellcolor[HTML]{cce6ff}3.91\% & \cellcolor[HTML]{EFFBEC}0.49\% \\
     & N@20 & & 0.0531 & 0.0724 & 0.0747 & 0.0660 & 0.0753 & 0.0727 & \underline{0.0777} & 0.0754 & \textbf{0.0779} & \cellcolor[HTML]{cce6ff}3.32\% & \cellcolor[HTML]{EFFBEC}0.26\% \\
     & N@50 & & 0.0770 & 0.0911 & 0.1016 & 0.0923 & 0.1019 & 0.1005 & \underline{0.1050} & 0.1024 & \textbf{0.1060} & \cellcolor[HTML]{cce6ff}3.52\% & \cellcolor[HTML]{EFFBEC}0.95\% \\
     
     & M@10 & & 0.0273 & 0.0428 & 0.0443 & 0.0382 & 0.0450 & 0.0431 & \underline{0.0461} & 0.0449 & \textbf{0.0463} & \cellcolor[HTML]{cce6ff}3.12\% & \cellcolor[HTML]{EFFBEC}0.43\% \\
     & M@20 & & 0.0310 & 0.0471 & 0.0487 & 0.0422 & 0.0494 & 0.0473 & \underline{\textbf{0.0508}} & 0.0494 & \textbf{0.0508} & \cellcolor[HTML]{cce6ff}2.83\% & \cellcolor[HTML]{fbfbea}0.00\%\\
     & M@50 & & 0.0347 & 0.0513 & 0.0529 & 0.0464 & 0.0536 & 0.0517 & \underline{\textbf{0.0552}} & 0.0536 & \textbf{0.0552} & \cellcolor[HTML]{cce6ff}2.99\% & \cellcolor[HTML]{fbfbea}0.00\%\\
  
    \hline
    \bottomrule
    \end{tabular}
    }
    }
    \label{tab:main_results}
  \end{table*}
  
  \subsubsection{\textbf{Evaluation Metrics}}
    We employ well-established recommendation performance metrics including Hit Ratio (\emph{HR@K}), Normalized Discounted Cumulative Gain (\emph{NDCG@K}), and Mean Reciprocal Rank (\emph{MRR@K}) as the criteria for experimental assessment. 
    In calculating these metrics, it is necessary to extract the top K scores from the model's final predicted ranking. We choose K values of 10, 20, and 50 to encompass short, medium, and long prediction lengths.
    
  \subsubsection{\textbf{Implementation Details}}
    Our experimental procedures adhere to the paradigm requirements set by the PyTorch \cite{Pytorch2019} and RecBole \cite{Recbole-1.2.0}. 
    All models in the experiment are implemented with a model dimension of 64, 
    which is the conventional choice for sequential recommendation models based on pure ID modeling. 
    For models based on the Mamba architecture, such as Mamba4Rec, SSD4Rec*, and TiM4Rec, the SSM state factor was consistently set to 32, the kernel size for 1D causal convolution was 4, and the block expansion factor for linear projections was 2.
    It is notable that the 1D causal convolutional layer within the Mamba architecture constrains the multi-head parameter to multiples of 8. However, in the context of sequential recommendation, this parameter typically holds values of 2 or 4 \cite{SASRec2018, BERT4Rec2019}. 
    Consequently, we modify the 1D causal convolutional layer to accommodate the specific modeling tasks of sequential recommendation, ultimately standardizing it to a value of 4.
    The backbone layers across all models are consistently configured to a depth of 2.
    To address the sparsity of the Amazon datasets, a dropout rate of 0.4 is used, compared to 0.2 for MovieLens-1M and KuaiRand. 
    The Adam optimizer \cite{Adam2015} is used uniformly with a learning rate set to 0.01. A batch size of 2048 is employed for training, and the validated batch size is 4096. 
    The maximum sequence length is set proportionally to the mean number of actions per user: 200 for MovieLens1M and 50 for Amazon-Beauty and KuaiRand datasets.
    We adhere to the optimal parameter configurations specific to each baseline, ensuring that the core parameters remain consistent across all models, and make adjustments to maximize performance to the extent feasible.

\subsection{Performance Comparison}
We initially conduct a comparative analysis of the recommendation performance between our proposed model and several baseline models across three benchmark datasets. 
The results of this comparison are presented in Table \ref{tab:main_results}. 
Through a detailed examination of the experimental outcomes, we are able to derive the following conclusions:
\begin{itemize}[leftmargin=10pt]
  \item In low-dimensional settings, the SSD architecture experiences a considerable decline in recommendation performance compared to the SSM architecture. However, when compared with models founded on Transformer architectures, it exhibits enhanced performance. 
  This observation aligns with our theoretical analysis presented in Section \ref{sec:mask-ssd}.
  \item TiM4Rec exhibits a significant performance enhancement when compared to SSD4Rec*, which is modeled on a purely SSD-based architecture, thereby confirming the efficacy of the Time-aware SSD Block. 
  Furthermore, TiM4Rec demonstrates a slight performance advantage over Mamba4Rec. 
  This observation underscores the effectiveness of leveraging time-difference in user interactions to capture shifts in user interests, 
  effectively mitigating the performance decline observed when transitioning from the SSM architecture to the SSD architecture.
\end{itemize}

\subsection{Dimension comparison}
To assess the suitability of modeling pure ID sequential recommendation tasks in lower dimensions, 
we compare the experimental results presented in the SSD4Rec paper at 256 dimensions. These results are shown in Table \ref{tab:256d_results}. 
Our research findings clearly demonstrate that the Mamba4Rec model tends to exhibit superior performance in 64-dimensional space. 
Additionally, the SSD4Rec* model that we replicate, which does not utilize variable-length sequences or bidirectional SSDs, 
showed only a slight decline in performance at 64 dimensions compared to the results presented in the original SSD4Rec paper at 256 dimensions. 
Notably, the performance of SSD4Rec at 256 dimensions is even inferior to that Mamba4Rec at 64 dimensions. 
These findings provide robust evidence that pure ID sequential recommendation tasks are more effectively modeled in lower-dimensional spaces. Moreover, 
the computational overhead associated with 256 dimensions is substantially higher, underscoring the advantages of utilizing our model. 
In addtional, a key advantage of the SSD architecture over the SSM architecture is its superior capability in high-dimensional modeling. 
To validate this advantage, we conduct a comparative analysis of the performance of our proposed method, TiM4Rec, at a dimensionality of 256. 
The results confirm that TiM4Rec effectively preserves the benefits of the SSD architecture even in high-dimensional settings. 
\begin{table}[h]
  \centering
  \caption{The performance of sequential recommendation models based on SSM and SSD is compared on the ML1M dataset in 64 and 256 dimensions. The definitions are aligned with the provisions of Table \ref{tab:main_results}.}
  \setlength{\tabcolsep}{0.2em}{
    \resizebox{0.5\textwidth}{!}{
      \begin{tabular}{clcccccc}
        \toprule
        \makecell{Dimension $\rightarrow$} &  & \multicolumn{3}{c}{64D} & \multicolumn{3}{c}{256D} \\
        \cmidrule(l){1-1} \cmidrule(l){3-5} \cmidrule(l){6-8}
        Metric & & {\small Mamba4Rec} & {\small SSD4Rec*}&{\small TiM4Rec} & {\small Mamba4Rec} & {\small SSD4Rec} & {\small TiM4Rec} \\
        \hline
        \midrule
        R@10 & & 0.3253 & 0.3199 & 0.3310 & 0.3124 & 0.3152 & 0.3270 \\
        R@20 & & 0.4354 & 0.4240 & 0.4338 & 0.4103 & 0.4194 & 0.4272 \\
        N@10 & & 0.1891 & 0.1841 & 0.1932 & 0.1847 & 0.1889 & 0.1938 \\
        N@20 & & 0.2169 & 0.2104 & 0.2194 & 0.2094 & 0.2145 & 0.2191 \\
        M@10 & & 0.1474 & 0.1425 & 0.1512 & 0.1456 & 0.1495 & 0.1529 \\
        M@20 & & 0.1551 & 0.1498 & 0.1584 & 0.1523 & 0.1565 & 0.1598 \\
        \hline
        \bottomrule
      \end{tabular}
    }
  }
  \label{tab:256d_results}
\end{table}
\begin{figure}[h]
  \centering
  \includegraphics[width=0.5\textwidth]{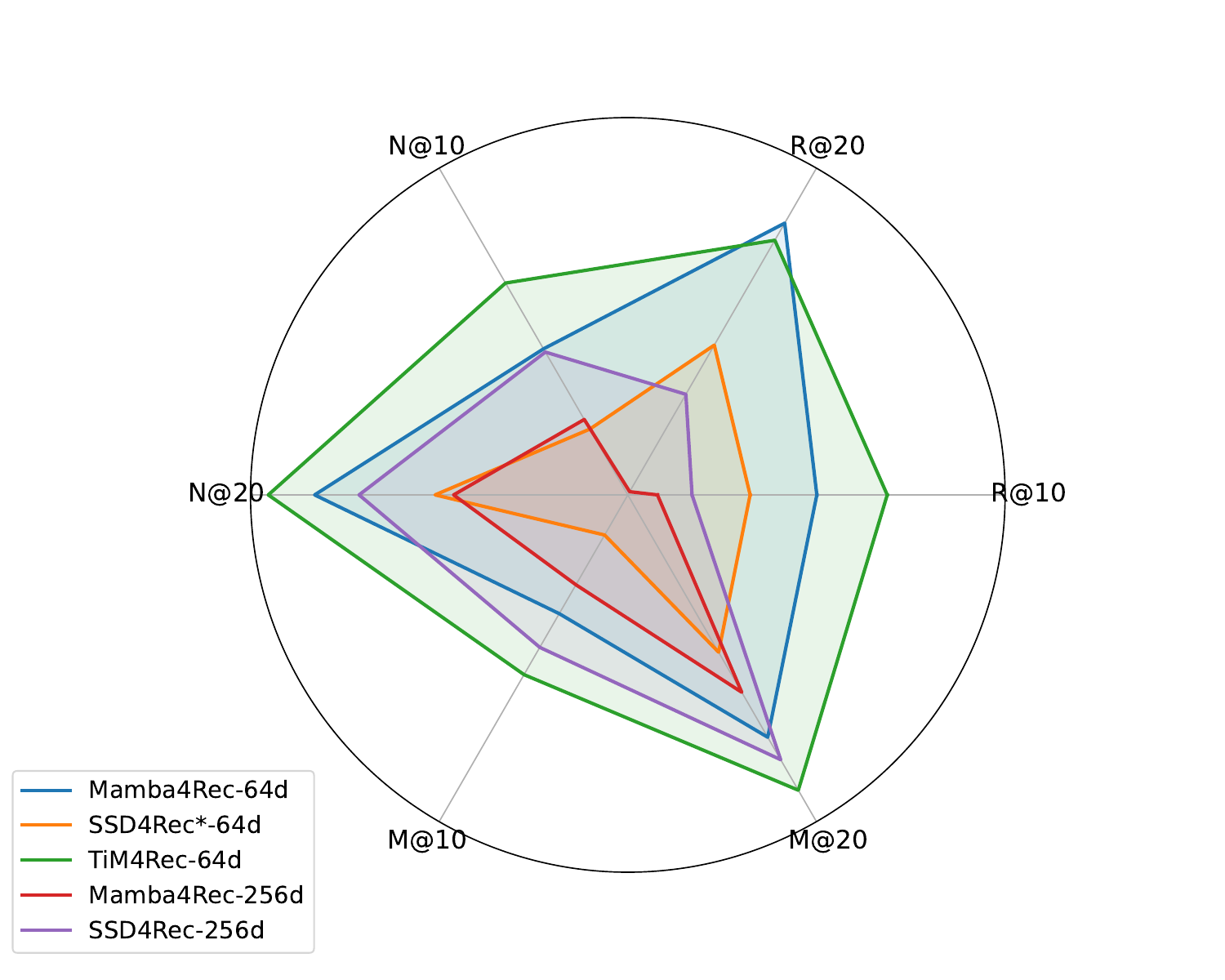}
  \caption{Regarding the performance radar chart of Table \ref{tab:256d_results}}
\end{figure}

\subsection{Efficiency Analysis}
In this section, we verify whether TiM4Rec can retain the inherent advantages of SSD architecture during model training and inference. 
Table \ref{tab:time_results} presents the training and inference times for representative models built on Transformer and Mamba architectures, all executed on a single NVIDIA RTX 3090 GPU.
\par
Observing the experimental results, we can draw the following conclusions:
\begin{itemize}[leftmargin=10pt]
    \item In both high-dimensional and low-dimensional spaces, TiM4Rec effectively preserves the inherent computational efficiency of the SSD architecture, necessitating only a minimal additional computational load when compared to SSD4Rec*, which is exclusively based on the SSD architecture. Notably, when compared with SASRec and Mamba4Rec, TiM4Rec achieves substantial enhancements in computational efficiency, attributed to leveraging the high-efficiency computing capabilities of the SSD architecture. This demonstrates TiM4Rec's ability to significantly optimize computational processes while minimizing overhead, thereby underscoring its advantage in scenarios demanding efficient and scalable solutions.
    \item Although TiSASRec maintains the computational complexity akin to that of SASRec, its time-aware methodology introduces notable inefficiencies, leading to a substantial escalation in actual computational load. By contrast, TiM4Rec demonstrates remarkable efficiency by necessitating only minimal additional computational resources across both high-dimensional and low-dimensional domains. The experimental results robustly validate the efficacy of the proposed time-aware enhancement method, which maintains linear computational complexity, underscoring its potential as an effective solution for efficient sequential recommendation systems.
\end{itemize}
\begin{table}[ht]
  \centering
  \caption{Comparison of training and inference times across various models using the ML-1M dataset. All models have a dimensionality of 64, except those with 256d suffix.}
  \setlength{\tabcolsep}{0.2em}{
    \resizebox{0.45\textwidth}{!}{
      \begin{tabular}{c|ccc}
        \toprule
        Model & Architecture & Training Time & Inference Time \\
        \hline
        \midrule
        SASRec & Attention & 207.42s & 0.81s \\
        TiSASRec & Attention & 1149.57s & 1.89s \\
        LRURec & LRU & 260.92s & 0.77s \\
        Mamba4Rec & SSM & 111.76s & 0.26s \\
        SSD4Rec* & SSD & 74.43s & 0.18s \\
        TiM4Rec & SSD & 83.57s & 0.21s \\
        \midrule
        \small{SSD4Rec*-256d} & SSD & 338.23s & 0.64s \\
        \small{TiM4Rec-256d} & SSD & 342.38s & 0.64s \\
        \hline
        \bottomrule
      \end{tabular}
    }
  }
  \label{tab:time_results}
\end{table}
\par
It is noteworthy that although we only compare the computational efficiency of our own replicated version of SSD4Rec*, the actual computational efficiency of SSD4Rec* surpasses that of SSD4Rec because SSD4Rec* eliminates the bidirectional SSD processing and the variable length sequence methods originally proposed in SSD4Rec, which come with additional overhead.
\par
\begin{figure}[h]
    \centering
    \begin{subfigure}{0.48\textwidth}
        \centering
        \includegraphics[width=1\textwidth]{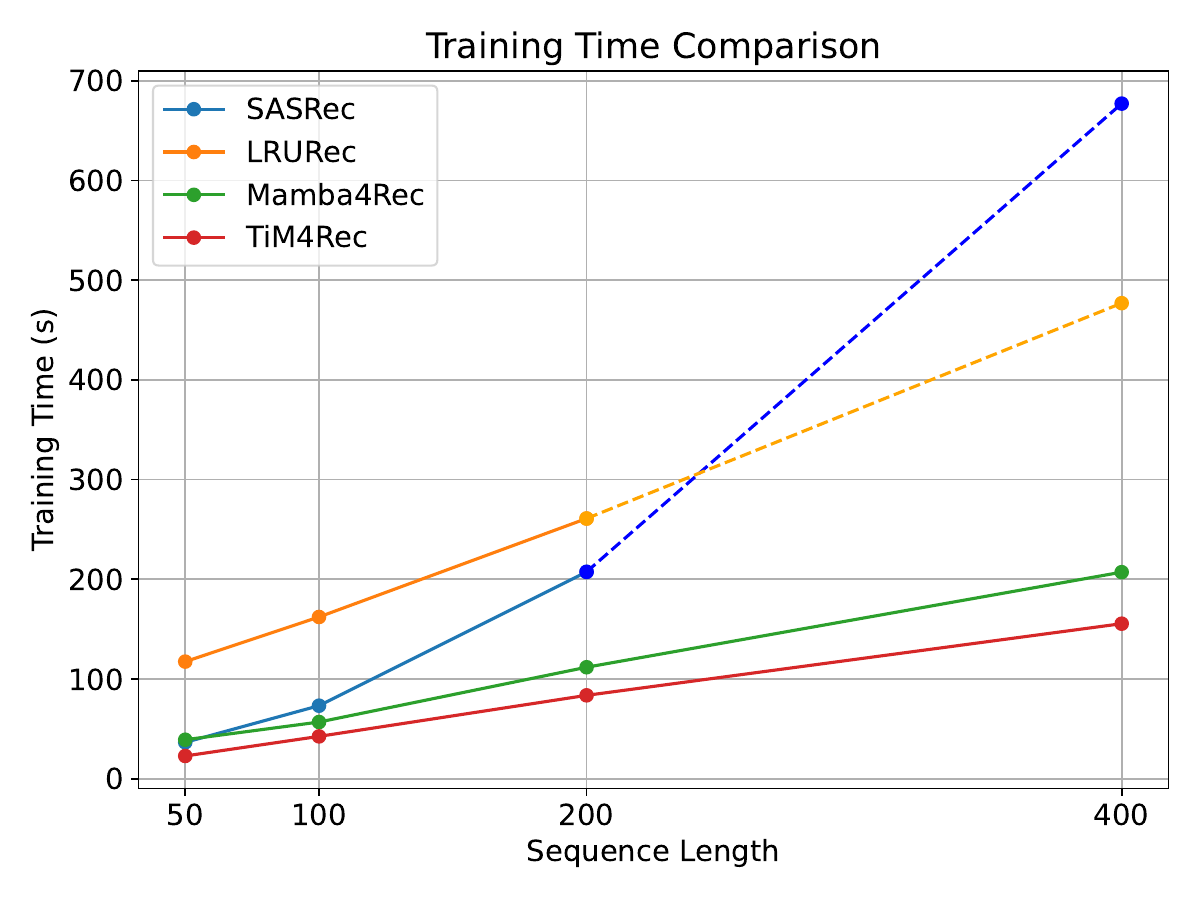}
        \caption{Training time comparison}
        \label{fig:train_time}
    \end{subfigure}
    \begin{subfigure}{0.48\textwidth}
        \centering
        \includegraphics[width=1\textwidth]{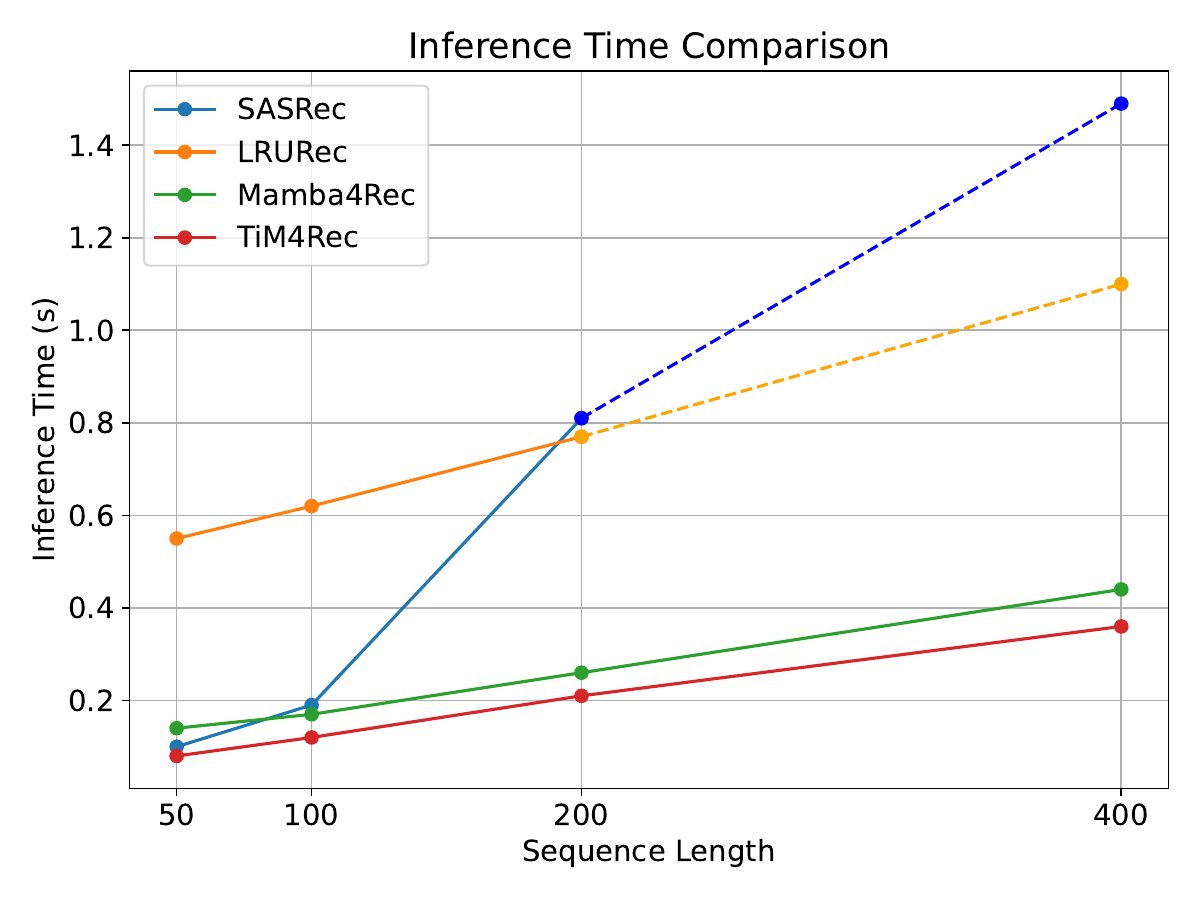}
        \caption{Inference time comparison}
        \label{fig:inference_time}
    \end{subfigure}
    \caption{Comparison of model training and inference times across different sequence lengths on ML-1M. The results for SASRec and LRURec at a sequence length of 400 were obtained by reducing the batch size due to Out-Of-Memory.}
    \label{fig:tran_and_inference_time}
\end{figure}

Additionally, we compared the training and inference times of various models across different sequence lengths, as shown in Figure \ref{fig:tran_and_inference_time}. The results indicate that TiM4Rec consistently maintains a significant speed advantage across all sequence lengths. Compared to the Transformer-based SASRec and the SSM-based Mamba4Rec, TiM4Rec effectively inherits the efficiency advantages of the SSD architecture while achieving superior performance. This result also demonstrates that our proposed time-aware enhancement method has linear computational complexity, ensuring that TiM4Rec does not experience efficiency degradation as sequence length increases.
\subsection{Ablation Study}
To validate the efficacy of our proposed time-aware algorithm along with other critical design elements, we perform a series of ablation studies using the ML-1M and Amazon-Beauty dataset. The procedure for the systematic elimination of each component is detailed as follows:
\begin{itemize}[leftmargin=10pt]
    \item w/o Time : Remove the proposed time-aware algorithm from the SSD module, rendering the model equivalent to our replication of the SSD4Rec* model mentioned earlier.
    \item w/o FFN : To assess the necessity of the FFN layer's nonlinear transformation ability in extracting sequential features, we eliminated this layer as part of our ablation study.
    \item N layer : Performance of the Model with N Stacked Layers of Time-aware SSD Blocks.
\end{itemize}
\par
Based on the results presented in Table \ref{tab:ablation-experiments}, it is evident that the time-aware enhancement method our propose significantly improves performance. The ablation experiment findings further underscore the critical role of the FFN layer's feature mapping ability in the extraction of sequence features. 
\begin{table*}[h]
    \centering
    \caption{Results of ablation studies}
    \setlength{\tabcolsep}{0.2em}{
      \resizebox{0.9\textwidth}{!}{
        \begin{tabular}{cc|cccccc|cccccc}
          \toprule
          Dataset $\rightarrow$ & & \multicolumn{6}{c}{ML-1M} & \multicolumn{6}{c}{Amazon-Beauty} \\
          \cmidrule(l){1-1} \cmidrule(l){3-8} \cmidrule(l){9-14}
          Component & & R@10 & R@20& N@10 & N@20 & M@10 & M@20 & R@10 & R@20& N@10 & N@20 & M@10 & M@20\\
          \hline
          \midrule
          Full & & \textbf{0.3310} & \textbf{0.4338} & \textbf{0.1932} & \textbf{0.2194} & \textbf{0.1512} & \textbf{0.1584} & \textbf{0.0854} & \underline{0.1204} & \textbf{0.0446} & \textbf{0.0533} & \underline{0.0321} & \underline{0.0345} \\
          w/o Time & & 0.3199 & 0.4240 & 0.1841 & 0.2104 & 0.1425 & 0.1498 & 0.0806 & 0.1146 & 0.0423 & 0.0509 & 0.0306 & 0.0329\\
          w/o FFN & & 0.3247 & 0.4278 & 0.1842 & 0.2103 & 0.1415 & 0.1487 & 0.0827 & 0.1183 & 0.0422 & 0.0512 & 0.0298 & 0.0322\\
          \midrule
          1 layer & & 0.3088 & 0.4146 & 0.1773 & 0.2040 & 0.1373 & 0.1446 & 0.0739 & 0.1042 & \underline{0.0437} & 0.0513 & \textbf{0.0344} & \textbf{0.0365}\\
          3 layer & & \underline{\textbf{0.3310}} & \underline{0.4323} & \underline{0.1916} & \underline{0.2172} & \underline{0.1490} & \underline{0.1560} & \underline{0.0848} & \textbf{0.1211} & 0.0431 & \underline{0.0522} & 0.0303 & 0.0328\\
          \hline
          \bottomrule
        \end{tabular}
      }
    }
      \label{tab:ablation-experiments}
\end{table*}

\section{Conclusion and Future Work}
% In this study, we first provide a detailed introduction and analysis of the SSM and SSD architectures. We then enhance the SSD architecture by introducing a time-aware mask matrix with linear computational complexity. This innovation addresses the performance degradation observed in SSD architectures when modeling in low-dimensional spaces. Furthermore, our research is the first to investigate the temporal enhancement techniques within the Mamba series architecture for sequence recommendation, achieving state-of-the-art performance without compromising the computational efficiency of SSDs. Looking ahead, we aim to explore the application of SSD architecture in the realm of multi-modal sequential recommendation, paving the way for sequential recommendation algorithms that better meet the demands of real-world recommendation systems.
In conclusion, this study makes significant strides in the advancement of sequential recommendation systems by thoroughly analyzing the SSM and SSD architectures. Our enhancements to the SSD framework, particularly through the integration of a time-aware mask matrix, effectively address the challenges associated with performance degradation in low-dimensional spaces. This novel approach not only preserves but also enhances the computational efficiency of SSDs, thereby setting a new benchmark in the field. Moreover, our research marks a pioneering effort in exploring temporal enhancement techniques within the Mamba series architecture, achieving remarkable state-of-the-art performance.
\par
Our contributions offer valuable insights into the development of efficient and robust recommendation systems, opening up new avenues for further exploration. Notably, our work emphasizes the importance of temporal dynamics in improving recommendation accuracy and efficiency. Moving forward, we are excited to extend our research into the domain of multi-modal sequential recommendation. By doing so, we aim to develop algorithms that are not only more aligned with the intricate demands of real-world recommendation systems but also capable of leveraging diverse data modalities for improved prediction and recommendation outcomes. This future direction holds promise for broadening the applicability and effectiveness of sequential recommendation algorithms, ultimately leading to more personalized and context-aware user experiences.

\begin{acks}
    The authors declare no potential conflict of interest. 
    We extend our gratitude to Master Guodong Sun from Northwestern Polytechnical University for his valuable revision suggestions for this article.
    The work was supported by Humanity and Social Science Foundation of Ministry of Education of China (No.18YJA630037, 21YJA630054), Zhejiang Province Soft Science Research Program Project (No.2024C350470).
\end{acks}

\balance
\bibliographystyle{ACM-Reference-Format}
\bibliography{ref}

\begin{thebibliography}{}

\end{thebibliography}



\begin{thebibliography}{54}

%%% the Web address.

\ifx \showCODEN    \undefined \def \showCODEN     #1{\unskip}     \fi
\ifx \showDOI      \undefined \def \showDOI       #1{#1}\fi
\ifx \showISBNx    \undefined \def \showISBNx     #1{\unskip}     \fi
\ifx \showISBNxiii \undefined \def \showISBNxiii  #1{\unskip}     \fi
\ifx \showISSN     \undefined \def \showISSN      #1{\unskip}     \fi
\ifx \showLCCN     \undefined \def \showLCCN      #1{\unskip}     \fi
\ifx \shownote     \undefined \def \shownote      #1{#1}          \fi
\ifx \showarticletitle \undefined \def \showarticletitle #1{#1}   \fi
\ifx \showURL      \undefined \def \showURL       {\relax}        \fi
% invisible to TeX
\providecommand\bibfield[2]{#2}
\providecommand\bibinfo[2]{#2}
\providecommand\natexlab[1]{#1}
\providecommand\showeprint[2][]{arXiv:#2}

\bibitem[Bontempelli et~al\mbox{.}(2022)]%
        {Deezer2022}
\bibfield{author}{\bibinfo{person}{Th{\'{e}}o Bontempelli}, \bibinfo{person}{Benjamin Chapus}, \bibinfo{person}{Fran{\c{c}}ois Rigaud}, \bibinfo{person}{Mathieu Morlon}, \bibinfo{person}{Marin Lorant}, {and} \bibinfo{person}{Guillaume Salha{-}Galvan}.} \bibinfo{year}{2022}\natexlab{}.
\newblock \showarticletitle{Flow Moods: Recommending Music by Moods on Deezer}. In \bibinfo{booktitle}{\emph{RecSys '22: Sixteenth {ACM} Conference on Recommender Systems, Seattle, WA, USA, September 18 - 23, 2022}}. \bibinfo{publisher}{{ACM}}, \bibinfo{pages}{452--455}.
\newblock
\urldef\tempurl%
\url{https://doi.org/10.1145/3523227.3547378}
\showDOI{\tempurl}


\bibitem[Chang et~al\mbox{.}(2021)]%
        {ChangGZHNSJ02021}
\bibfield{author}{\bibinfo{person}{Jianxin Chang}, \bibinfo{person}{Chen Gao}, \bibinfo{person}{Yu Zheng}, \bibinfo{person}{Yiqun Hui}, \bibinfo{person}{Yanan Niu}, \bibinfo{person}{Yang Song}, \bibinfo{person}{Depeng Jin}, {and} \bibinfo{person}{Yong Li}.} \bibinfo{year}{2021}\natexlab{}.
\newblock \showarticletitle{Sequential Recommendation with Graph Neural Networks}. In \bibinfo{booktitle}{\emph{{SIGIR} '21: The 44th International {ACM} {SIGIR} Conference on Research and Development in Information Retrieval, Virtual Event, Canada, July 11-15, 2021}}. \bibinfo{publisher}{{ACM}}, \bibinfo{pages}{378--387}.
\newblock
\urldef\tempurl%
\url{https://doi.org/10.1145/3404835.3462968}
\showDOI{\tempurl}


\bibitem[Chen et~al\mbox{.}(2024)]%
        {BTMT2024}
\bibfield{author}{\bibinfo{person}{Ruizhen Chen}, \bibinfo{person}{Yihao Zhang}, \bibinfo{person}{Jiahao Hu}, \bibinfo{person}{Xibin Wang}, \bibinfo{person}{Junlin Zhu}, {and} \bibinfo{person}{Weiwen Liao}.} \bibinfo{year}{2024}\natexlab{}.
\newblock \showarticletitle{Behavior sessions and time-aware for multi-target sequential recommendation}.
\newblock \bibinfo{journal}{\emph{Appl. Intell.}} \bibinfo{volume}{54}, \bibinfo{number}{20} (\bibinfo{year}{2024}), \bibinfo{pages}{9830--9847}.
\newblock
\urldef\tempurl%
\url{https://doi.org/10.1007/S10489-024-05678-6}
\showDOI{\tempurl}


\bibitem[Cho et~al\mbox{.}(2021)]%
        {TimelyRec2021}
\bibfield{author}{\bibinfo{person}{Junsu Cho}, \bibinfo{person}{Dongmin Hyun}, \bibinfo{person}{SeongKu Kang}, {and} \bibinfo{person}{Hwanjo Yu}.} \bibinfo{year}{2021}\natexlab{}.
\newblock \showarticletitle{Learning Heterogeneous Temporal Patterns of User Preference for Timely Recommendation}. In \bibinfo{booktitle}{\emph{{WWW} '21: The Web Conference 2021, Virtual Event / Ljubljana, Slovenia, April 19-23, 2021}}. \bibinfo{publisher}{{ACM} / {IW3C2}}, \bibinfo{pages}{1274--1283}.
\newblock
\urldef\tempurl%
\url{https://doi.org/10.1145/3442381.3449947}
\showDOI{\tempurl}


\bibitem[Cho et~al\mbox{.}(2020)]%
        {MEANTIME2020}
\bibfield{author}{\bibinfo{person}{Sung~Min Cho}, \bibinfo{person}{Eunhyeok Park}, {and} \bibinfo{person}{Sungjoo Yoo}.} \bibinfo{year}{2020}\natexlab{}.
\newblock \showarticletitle{{MEANTIME:} Mixture of Attention Mechanisms with Multi-temporal Embeddings for Sequential Recommendation}. In \bibinfo{booktitle}{\emph{RecSys 2020: Fourteenth {ACM} Conference on Recommender Systems, Virtual Event, Brazil, September 22-26, 2020}}. \bibinfo{publisher}{{ACM}}, \bibinfo{pages}{515--520}.
\newblock
\urldef\tempurl%
\url{https://doi.org/10.1145/3383313.3412216}
\showDOI{\tempurl}


\bibitem[Dang et~al\mbox{.}(2023)]%
        {TiCoSeRec2023}
\bibfield{author}{\bibinfo{person}{Yizhou Dang}, \bibinfo{person}{Enneng Yang}, \bibinfo{person}{Guibing Guo}, \bibinfo{person}{Linying Jiang}, \bibinfo{person}{Xingwei Wang}, \bibinfo{person}{Xiaoxiao Xu}, \bibinfo{person}{Qinghui Sun}, {and} \bibinfo{person}{Hong Liu}.} \bibinfo{year}{2023}\natexlab{}.
\newblock \showarticletitle{Uniform Sequence Better: Time Interval Aware Data Augmentation for Sequential Recommendation}. In \bibinfo{booktitle}{\emph{Thirty-Seventh {AAAI} Conference on Artificial Intelligence, {AAAI} 2023, Thirty-Fifth Conference on Innovative Applications of Artificial Intelligence, {IAAI} 2023, Thirteenth Symposium on Educational Advances in Artificial Intelligence, {EAAI} 2023, Washington, DC, USA, February 7-14, 2023}}. \bibinfo{publisher}{{AAAI} Press}, \bibinfo{pages}{4225--4232}.
\newblock
\urldef\tempurl%
\url{https://doi.org/10.1609/AAAI.V37I4.25540}
\showDOI{\tempurl}


\bibitem[Dao et~al\mbox{.}(2022)]%
        {FlashAttention2022}
\bibfield{author}{\bibinfo{person}{Tri Dao}, \bibinfo{person}{Daniel~Y. Fu}, \bibinfo{person}{Stefano Ermon}, \bibinfo{person}{Atri Rudra}, {and} \bibinfo{person}{Christopher R{\'{e}}}.} \bibinfo{year}{2022}\natexlab{}.
\newblock \showarticletitle{FlashAttention: Fast and Memory-Efficient Exact Attention with IO-Awareness}. In \bibinfo{booktitle}{\emph{Advances in Neural Information Processing Systems 35: Annual Conference on Neural Information Processing Systems 2022, NeurIPS 2022, New Orleans, LA, USA, November 28 - December 9, 2022}}. \bibinfo{pages}{1--34}.
\newblock
\urldef\tempurl%
\url{http://papers.nips.cc/paper\_files/paper/2022/hash/67d57c32e20fd0a7a302cb81d36e40d5-Abstract-Conference.html}
\showURL{%
\tempurl}


\bibitem[Dao and Gu(2024)]%
        {SSD2024}
\bibfield{author}{\bibinfo{person}{Tri Dao} {and} \bibinfo{person}{Albert Gu}.} \bibinfo{year}{2024}\natexlab{}.
\newblock \showarticletitle{Transformers are SSMs: Generalized Models and Efficient Algorithms Through Structured State Space Duality}. In \bibinfo{booktitle}{\emph{Forty-first International Conference on Machine Learning, {ICML} 2024, Vienna, Austria, July 21-27, 2024}}. \bibinfo{publisher}{OpenReview.net}, \bibinfo{pages}{1--31}.
\newblock
\urldef\tempurl%
\url{https://openreview.net/forum?id=ztn8FCR1td}
\showURL{%
\tempurl}


\bibitem[Devlin et~al\mbox{.}(2019)]%
        {BERT2019}
\bibfield{author}{\bibinfo{person}{Jacob Devlin}, \bibinfo{person}{Ming{-}Wei Chang}, \bibinfo{person}{Kenton Lee}, {and} \bibinfo{person}{Kristina Toutanova}.} \bibinfo{year}{2019}\natexlab{}.
\newblock \showarticletitle{{BERT:} Pre-training of Deep Bidirectional Transformers for Language Understanding}. In \bibinfo{booktitle}{\emph{Proceedings of the 2019 Conference of the North American Chapter of the Association for Computational Linguistics: Human Language Technologies, {NAACL-HLT} 2019, Minneapolis, MN, USA, June 2-7, 2019, Volume 1 (Long and Short Papers)}}. \bibinfo{publisher}{Association for Computational Linguistics}, \bibinfo{pages}{4171--4186}.
\newblock
\urldef\tempurl%
\url{https://doi.org/10.18653/V1/N19-1423}
\showDOI{\tempurl}


\bibitem[Gao et~al\mbox{.}(2022)]%
        {KuaiRand2022}
\bibfield{author}{\bibinfo{person}{Chongming Gao}, \bibinfo{person}{Shijun Li}, \bibinfo{person}{Yuan Zhang}, \bibinfo{person}{Jiawei Chen}, \bibinfo{person}{Biao Li}, \bibinfo{person}{Wenqiang Lei}, \bibinfo{person}{Peng Jiang}, {and} \bibinfo{person}{Xiangnan He}.} \bibinfo{year}{2022}\natexlab{}.
\newblock \showarticletitle{KuaiRand: An Unbiased Sequential Recommendation Dataset with Randomly Exposed Videos}. In \bibinfo{booktitle}{\emph{Proceedings of the 31st {ACM} International Conference on Information {\&} Knowledge Management, Atlanta, GA, USA, October 17-21, 2022}}. \bibinfo{publisher}{{ACM}}, \bibinfo{pages}{3953--3957}.
\newblock
\urldef\tempurl%
\url{https://doi.org/10.1145/3511808.3557624}
\showDOI{\tempurl}


\bibitem[Gu and Dao(2023)]%
        {Mamba2023}
\bibfield{author}{\bibinfo{person}{Albert Gu} {and} \bibinfo{person}{Tri Dao}.} \bibinfo{year}{2023}\natexlab{}.
\newblock \showarticletitle{Mamba: Linear-Time Sequence Modeling with Selective State Spaces}.
\newblock \bibinfo{journal}{\emph{CoRR}}  \bibinfo{volume}{abs/2312.00752} (\bibinfo{year}{2023}).
\newblock
\urldef\tempurl%
\url{https://doi.org/10.48550/ARXIV.2312.00752}
\showDOI{\tempurl}
\showeprint[arXiv]{2312.00752}


\bibitem[Gu et~al\mbox{.}(2020)]%
        {HiPPO2020}
\bibfield{author}{\bibinfo{person}{Albert Gu}, \bibinfo{person}{Tri Dao}, \bibinfo{person}{Stefano Ermon}, \bibinfo{person}{Atri Rudra}, {and} \bibinfo{person}{Christopher R{\'{e}}}.} \bibinfo{year}{2020}\natexlab{}.
\newblock \showarticletitle{HiPPO: Recurrent Memory with Optimal Polynomial Projections}. In \bibinfo{booktitle}{\emph{Advances in Neural Information Processing Systems 33: Annual Conference on Neural Information Processing Systems 2020, NeurIPS 2020, December 6-12, 2020, virtual}}. \bibinfo{pages}{1--14}.
\newblock
\urldef\tempurl%
\url{https://proceedings.neurips.cc/paper/2020/hash/102f0bb6efb3a6128a3c750dd16729be-Abstract.html}
\showURL{%
\tempurl}


\bibitem[Gu et~al\mbox{.}(2022)]%
        {S4-2022}
\bibfield{author}{\bibinfo{person}{Albert Gu}, \bibinfo{person}{Karan Goel}, {and} \bibinfo{person}{Christopher R{\'{e}}}.} \bibinfo{year}{2022}\natexlab{}.
\newblock \showarticletitle{Efficiently Modeling Long Sequences with Structured State Spaces}. In \bibinfo{booktitle}{\emph{The Tenth International Conference on Learning Representations, {ICLR} 2022, Virtual Event, April 25-29, 2022}}. \bibinfo{publisher}{OpenReview.net}, \bibinfo{pages}{1--27}.
\newblock
\urldef\tempurl%
\url{https://openreview.net/forum?id=uYLFoz1vlAC}
\showURL{%
\tempurl}


\bibitem[Gu et~al\mbox{.}(2021)]%
        {SSM2021}
\bibfield{author}{\bibinfo{person}{Albert Gu}, \bibinfo{person}{Isys Johnson}, \bibinfo{person}{Karan Goel}, \bibinfo{person}{Khaled Saab}, \bibinfo{person}{Tri Dao}, \bibinfo{person}{Atri Rudra}, {and} \bibinfo{person}{Christopher R{\'{e}}}.} \bibinfo{year}{2021}\natexlab{}.
\newblock \showarticletitle{Combining Recurrent, Convolutional, and Continuous-time Models with Linear State Space Layers}. In \bibinfo{booktitle}{\emph{Advances in Neural Information Processing Systems 34: Annual Conference on Neural Information Processing Systems 2021, NeurIPS 2021, December 6-14, 2021, virtual}}, \bibfield{editor}{\bibinfo{person}{Marc'Aurelio Ranzato}, \bibinfo{person}{Alina Beygelzimer}, \bibinfo{person}{Yann~N. Dauphin}, \bibinfo{person}{Percy Liang}, {and} \bibinfo{person}{Jennifer~Wortman Vaughan}} (Eds.). \bibinfo{pages}{572--585}.
\newblock
\urldef\tempurl%
\url{https://proceedings.neurips.cc/paper/2021/hash/05546b0e38ab9175cd905eebcc6ebb76-Abstract.html}
\showURL{%
\tempurl}


\bibitem[Hansen et~al\mbox{.}(2020)]%
        {MusicRec2020}
\bibfield{author}{\bibinfo{person}{Casper Hansen}, \bibinfo{person}{Christian Hansen}, \bibinfo{person}{Lucas Maystre}, \bibinfo{person}{Rishabh Mehrotra}, \bibinfo{person}{Brian Brost}, \bibinfo{person}{Federico Tomasi}, {and} \bibinfo{person}{Mounia Lalmas}.} \bibinfo{year}{2020}\natexlab{}.
\newblock \showarticletitle{Contextual and Sequential User Embeddings for Large-Scale Music Recommendation}. In \bibinfo{booktitle}{\emph{RecSys 2020: Fourteenth {ACM} Conference on Recommender Systems, Virtual Event, Brazil, September 22-26, 2020}}. \bibinfo{publisher}{{ACM}}, \bibinfo{pages}{53--62}.
\newblock
\urldef\tempurl%
\url{https://doi.org/10.1145/3383313.3412248}
\showDOI{\tempurl}


\bibitem[Harper and Konstan(2016)]%
        {MovieLes2016}
\bibfield{author}{\bibinfo{person}{F.~Maxwell Harper} {and} \bibinfo{person}{Joseph~A. Konstan}.} \bibinfo{year}{2016}\natexlab{}.
\newblock \showarticletitle{The MovieLens Datasets: History and Context}.
\newblock \bibinfo{journal}{\emph{{ACM} Trans. Interact. Intell. Syst.}} \bibinfo{volume}{5}, \bibinfo{number}{4} (\bibinfo{year}{2016}), \bibinfo{pages}{19:1--19:19}.
\newblock
\urldef\tempurl%
\url{https://doi.org/10.1145/2827872}
\showDOI{\tempurl}


\bibitem[He et~al\mbox{.}(2016)]%
        {ResNet2016}
\bibfield{author}{\bibinfo{person}{Kaiming He}, \bibinfo{person}{Xiangyu Zhang}, \bibinfo{person}{Shaoqing Ren}, {and} \bibinfo{person}{Jian Sun}.} \bibinfo{year}{2016}\natexlab{}.
\newblock \showarticletitle{Deep Residual Learning for Image Recognition}. In \bibinfo{booktitle}{\emph{2016 {IEEE} Conference on Computer Vision and Pattern Recognition, {CVPR} 2016, Las Vegas, NV, USA, June 27-30, 2016}}. \bibinfo{publisher}{{IEEE} Computer Society}, \bibinfo{pages}{770--778}.
\newblock
\urldef\tempurl%
\url{https://doi.org/10.1109/CVPR.2016.90}
\showDOI{\tempurl}


\bibitem[He and McAuley(2016)]%
        {VBPR2016}
\bibfield{author}{\bibinfo{person}{Ruining He} {and} \bibinfo{person}{Julian~J. McAuley}.} \bibinfo{year}{2016}\natexlab{}.
\newblock \showarticletitle{{VBPR:} Visual Bayesian Personalized Ranking from Implicit Feedback}. In \bibinfo{booktitle}{\emph{Proceedings of the Thirtieth {AAAI} Conference on Artificial Intelligence, February 12-17, 2016, Phoenix, Arizona, {USA}}}. \bibinfo{publisher}{{AAAI} Press}, \bibinfo{pages}{144--150}.
\newblock
\urldef\tempurl%
\url{https://doi.org/10.1609/AAAI.V30I1.9973}
\showDOI{\tempurl}


\bibitem[Hidasi et~al\mbox{.}(2016)]%
        {GRU4Rec2016}
\bibfield{author}{\bibinfo{person}{Bal{\'{a}}zs Hidasi}, \bibinfo{person}{Alexandros Karatzoglou}, \bibinfo{person}{Linas Baltrunas}, {and} \bibinfo{person}{Domonkos Tikk}.} \bibinfo{year}{2016}\natexlab{}.
\newblock \showarticletitle{Session-based Recommendations with Recurrent Neural Networks}. In \bibinfo{booktitle}{\emph{4th International Conference on Learning Representations, {ICLR} 2016, San Juan, Puerto Rico, May 2-4, 2016, Conference Track Proceedings}}. \bibinfo{pages}{1--10}.
\newblock
\urldef\tempurl%
\url{http://arxiv.org/abs/1511.06939}
\showURL{%
\tempurl}


\bibitem[Hidasi and Tikk(2016)]%
        {HRM2016}
\bibfield{author}{\bibinfo{person}{Bal{\'{a}}zs Hidasi} {and} \bibinfo{person}{Domonkos Tikk}.} \bibinfo{year}{2016}\natexlab{}.
\newblock \showarticletitle{General factorization framework for context-aware recommendations}.
\newblock \bibinfo{journal}{\emph{Data Min. Knowl. Discov.}} \bibinfo{volume}{30}, \bibinfo{number}{2} (\bibinfo{year}{2016}), \bibinfo{pages}{342--371}.
\newblock
\urldef\tempurl%
\url{https://doi.org/10.1007/S10618-015-0417-Y}
\showDOI{\tempurl}


\bibitem[Huang et~al\mbox{.}(2015)]%
        {TencentRec2015}
\bibfield{author}{\bibinfo{person}{Yanxiang Huang}, \bibinfo{person}{Bin Cui}, \bibinfo{person}{Wenyu Zhang}, \bibinfo{person}{Jie Jiang}, {and} \bibinfo{person}{Ying Xu}.} \bibinfo{year}{2015}\natexlab{}.
\newblock \showarticletitle{TencentRec: Real-time Stream Recommendation in Practice}. In \bibinfo{booktitle}{\emph{Proceedings of the 2015 {ACM} {SIGMOD} International Conference on Management of Data, Melbourne, Victoria, Australia, May 31 - June 4, 2015}}. \bibinfo{publisher}{{ACM}}, \bibinfo{pages}{227--238}.
\newblock
\urldef\tempurl%
\url{https://doi.org/10.1145/2723372.2742785}
\showDOI{\tempurl}


\bibitem[Kang and McAuley(2018)]%
        {SASRec2018}
\bibfield{author}{\bibinfo{person}{Wang{-}Cheng Kang} {and} \bibinfo{person}{Julian~J. McAuley}.} \bibinfo{year}{2018}\natexlab{}.
\newblock \showarticletitle{Self-Attentive Sequential Recommendation}. In \bibinfo{booktitle}{\emph{{IEEE} International Conference on Data Mining, {ICDM} 2018, Singapore, November 17-20, 2018}}. \bibinfo{publisher}{{IEEE} Computer Society}, \bibinfo{pages}{197--206}.
\newblock
\urldef\tempurl%
\url{https://doi.org/10.1109/ICDM.2018.00035}
\showDOI{\tempurl}


\bibitem[Katharopoulos et~al\mbox{.}(2020)]%
        {LinearAttention2020}
\bibfield{author}{\bibinfo{person}{Angelos Katharopoulos}, \bibinfo{person}{Apoorv Vyas}, \bibinfo{person}{Nikolaos Pappas}, {and} \bibinfo{person}{Fran{\c{c}}ois Fleuret}.} \bibinfo{year}{2020}\natexlab{}.
\newblock \showarticletitle{Transformers are RNNs: Fast Autoregressive Transformers with Linear Attention}. In \bibinfo{booktitle}{\emph{Proceedings of the 37th International Conference on Machine Learning, {ICML} 2020, 13-18 July 2020, Virtual Event}} \emph{(\bibinfo{series}{Proceedings of Machine Learning Research}, Vol.~\bibinfo{volume}{119})}. \bibinfo{publisher}{{PMLR}}, \bibinfo{pages}{5156--5165}.
\newblock
\urldef\tempurl%
\url{http://proceedings.mlr.press/v119/katharopoulos20a.html}
\showURL{%
\tempurl}


\bibitem[Kingma and Ba(2015)]%
        {Adam2015}
\bibfield{author}{\bibinfo{person}{Diederik~P. Kingma} {and} \bibinfo{person}{Jimmy Ba}.} \bibinfo{year}{2015}\natexlab{}.
\newblock \showarticletitle{Adam: {A} Method for Stochastic Optimization}. In \bibinfo{booktitle}{\emph{3rd International Conference on Learning Representations, {ICLR} 2015, San Diego, CA, USA, May 7-9, 2015, Conference Track Proceedings}}. \bibinfo{pages}{1--15}.
\newblock
\urldef\tempurl%
\url{http://arxiv.org/abs/1412.6980}
\showURL{%
\tempurl}


\bibitem[Li et~al\mbox{.}(2017)]%
        {LiRCRLM2017}
\bibfield{author}{\bibinfo{person}{Jing Li}, \bibinfo{person}{Pengjie Ren}, \bibinfo{person}{Zhumin Chen}, \bibinfo{person}{Zhaochun Ren}, \bibinfo{person}{Tao Lian}, {and} \bibinfo{person}{Jun Ma}.} \bibinfo{year}{2017}\natexlab{}.
\newblock \showarticletitle{Neural Attentive Session-based Recommendation}. In \bibinfo{booktitle}{\emph{Proceedings of the 2017 {ACM} on Conference on Information and Knowledge Management, {CIKM} 2017, Singapore, November 06 - 10, 2017}}. \bibinfo{publisher}{{ACM}}, \bibinfo{pages}{1419--1428}.
\newblock
\urldef\tempurl%
\url{https://doi.org/10.1145/3132847.3132926}
\showDOI{\tempurl}


\bibitem[Li et~al\mbox{.}(2020)]%
        {TiSASRec2020}
\bibfield{author}{\bibinfo{person}{Jiacheng Li}, \bibinfo{person}{Yujie Wang}, {and} \bibinfo{person}{Julian~J. McAuley}.} \bibinfo{year}{2020}\natexlab{}.
\newblock \showarticletitle{Time Interval Aware Self-Attention for Sequential Recommendation}. In \bibinfo{booktitle}{\emph{{WSDM} '20: The Thirteenth {ACM} International Conference on Web Search and Data Mining, Houston, TX, USA, February 3-7, 2020}}. \bibinfo{publisher}{{ACM}}, \bibinfo{pages}{322--330}.
\newblock
\urldef\tempurl%
\url{https://doi.org/10.1145/3336191.3371786}
\showDOI{\tempurl}


\bibitem[Liu et~al\mbox{.}(2024a)]%
        {RecBLR2024}
\bibfield{author}{\bibinfo{person}{Chengkai Liu}, \bibinfo{person}{Jianghao Lin}, \bibinfo{person}{Hanzhou Liu}, \bibinfo{person}{Jianling Wang}, {and} \bibinfo{person}{James Caverlee}.} \bibinfo{year}{2024}\natexlab{a}.
\newblock \showarticletitle{Behavior-Dependent Linear Recurrent Units for Efficient Sequential Recommendation}. In \bibinfo{booktitle}{\emph{Proceedings of the 33rd {ACM} International Conference on Information and Knowledge Management, {CIKM} 2024, Boise, ID, USA, October 21-25, 2024}}. \bibinfo{publisher}{{ACM}}, \bibinfo{pages}{1430--1440}.
\newblock
\urldef\tempurl%
\url{https://doi.org/10.1145/3627673.3679717}
\showDOI{\tempurl}


\bibitem[Liu et~al\mbox{.}(2024b)]%
        {Mamba4Rec2024}
\bibfield{author}{\bibinfo{person}{Chengkai Liu}, \bibinfo{person}{Jianghao Lin}, \bibinfo{person}{Jianling Wang}, \bibinfo{person}{Hanzhou Liu}, {and} \bibinfo{person}{James Caverlee}.} \bibinfo{year}{2024}\natexlab{b}.
\newblock \showarticletitle{Mamba4Rec: Towards Efficient Sequential Recommendation with Selective State Space Models}.
\newblock \bibinfo{journal}{\emph{CoRR}}  \bibinfo{volume}{abs/2403.03900} (\bibinfo{year}{2024}).
\newblock
\urldef\tempurl%
\url{https://doi.org/10.48550/ARXIV.2403.03900}
\showDOI{\tempurl}
\showeprint[arXiv]{2403.03900}


\bibitem[Lu et~al\mbox{.}(2015)]%
        {RSSurvey2005}
\bibfield{author}{\bibinfo{person}{Jie Lu}, \bibinfo{person}{Dianshuang Wu}, \bibinfo{person}{Mingsong Mao}, \bibinfo{person}{Wei Wang}, {and} \bibinfo{person}{Guangquan Zhang}.} \bibinfo{year}{2015}\natexlab{}.
\newblock \showarticletitle{Recommender system application developments: {A} survey}.
\newblock \bibinfo{journal}{\emph{Decis. Support Syst.}}  \bibinfo{volume}{74} (\bibinfo{year}{2015}), \bibinfo{pages}{12--32}.
\newblock
\urldef\tempurl%
\url{https://doi.org/10.1016/J.DSS.2015.03.008}
\showDOI{\tempurl}


\bibitem[Ma et~al\mbox{.}(2020)]%
        {MaNLD2020}
\bibfield{author}{\bibinfo{person}{Yifei Ma}, \bibinfo{person}{Balakrishnan~(Murali) Narayanaswamy}, \bibinfo{person}{Haibin Lin}, {and} \bibinfo{person}{Hao Ding}.} \bibinfo{year}{2020}\natexlab{}.
\newblock \showarticletitle{Temporal-Contextual Recommendation in Real-Time}. In \bibinfo{booktitle}{\emph{{KDD} '20: The 26th {ACM} {SIGKDD} Conference on Knowledge Discovery and Data Mining, Virtual Event, CA, USA, August 23-27, 2020}}. \bibinfo{publisher}{{ACM}}, \bibinfo{pages}{2291--2299}.
\newblock
\urldef\tempurl%
\url{https://doi.org/10.1145/3394486.3403278}
\showDOI{\tempurl}


\bibitem[Martin and Cundy(2018)]%
        {Parallelizing2018}
\bibfield{author}{\bibinfo{person}{Eric Martin} {and} \bibinfo{person}{Chris Cundy}.} \bibinfo{year}{2018}\natexlab{}.
\newblock \showarticletitle{Parallelizing Linear Recurrent Neural Nets Over Sequence Length}. In \bibinfo{booktitle}{\emph{6th International Conference on Learning Representations, {ICLR} 2018, Vancouver, BC, Canada, April 30 - May 3, 2018, Conference Track Proceedings}}. \bibinfo{publisher}{OpenReview.net}, \bibinfo{pages}{1--9}.
\newblock
\urldef\tempurl%
\url{https://openreview.net/forum?id=HyUNwulC-}
\showURL{%
\tempurl}


\bibitem[McAuley et~al\mbox{.}(2015)]%
        {Amazon2014}
\bibfield{author}{\bibinfo{person}{Julian~J. McAuley}, \bibinfo{person}{Christopher Targett}, \bibinfo{person}{Qinfeng Shi}, {and} \bibinfo{person}{Anton van~den Hengel}.} \bibinfo{year}{2015}\natexlab{}.
\newblock \showarticletitle{Image-Based Recommendations on Styles and Substitutes}. In \bibinfo{booktitle}{\emph{Proceedings of the 38th International {ACM} {SIGIR} Conference on Research and Development in Information Retrieval, Santiago, Chile, August 9-13, 2015}}. \bibinfo{publisher}{{ACM}}, \bibinfo{pages}{43--52}.
\newblock
\urldef\tempurl%
\url{https://doi.org/10.1145/2766462.2767755}
\showDOI{\tempurl}


\bibitem[Orvieto et~al\mbox{.}(2023)]%
        {LRU2023}
\bibfield{author}{\bibinfo{person}{Antonio Orvieto}, \bibinfo{person}{Samuel~L. Smith}, \bibinfo{person}{Albert Gu}, \bibinfo{person}{Anushan Fernando}, \bibinfo{person}{{\c{C}}aglar G{\"{u}}l{\c{c}}ehre}, \bibinfo{person}{Razvan Pascanu}, {and} \bibinfo{person}{Soham De}.} \bibinfo{year}{2023}\natexlab{}.
\newblock \showarticletitle{Resurrecting Recurrent Neural Networks for Long Sequences}. In \bibinfo{booktitle}{\emph{International Conference on Machine Learning, {ICML} 2023, 23-29 July 2023, Honolulu, Hawaii, {USA}}} \emph{(\bibinfo{series}{Proceedings of Machine Learning Research}, Vol.~\bibinfo{volume}{202})}. \bibinfo{publisher}{{PMLR}}, \bibinfo{pages}{26670--26698}.
\newblock
\urldef\tempurl%
\url{https://proceedings.mlr.press/v202/orvieto23a.html}
\showURL{%
\tempurl}


\bibitem[Park et~al\mbox{.}(2024)]%
        {MambaNLP2024}
\bibfield{author}{\bibinfo{person}{Jongho Park}, \bibinfo{person}{Jaeseung Park}, \bibinfo{person}{Zheyang Xiong}, \bibinfo{person}{Nayoung Lee}, \bibinfo{person}{Jaewoong Cho}, \bibinfo{person}{Samet Oymak}, \bibinfo{person}{Kangwook Lee}, {and} \bibinfo{person}{Dimitris Papailiopoulos}.} \bibinfo{year}{2024}\natexlab{}.
\newblock \showarticletitle{Can Mamba Learn How To Learn? {A} Comparative Study on In-Context Learning Tasks}. In \bibinfo{booktitle}{\emph{Forty-first International Conference on Machine Learning, {ICML} 2024, Vienna, Austria, July 21-27, 2024}}. \bibinfo{publisher}{OpenReview.net}, \bibinfo{pages}{1--20}.
\newblock
\urldef\tempurl%
\url{https://openreview.net/forum?id=GbFluKMmtE}
\showURL{%
\tempurl}


\bibitem[Paszke et~al\mbox{.}(2019)]%
        {Pytorch2019}
\bibfield{author}{\bibinfo{person}{Adam Paszke}, \bibinfo{person}{Sam Gross}, \bibinfo{person}{Francisco Massa}, \bibinfo{person}{Adam Lerer}, \bibinfo{person}{James Bradbury}, \bibinfo{person}{Gregory Chanan}, \bibinfo{person}{Trevor Killeen}, \bibinfo{person}{Zeming Lin}, \bibinfo{person}{Natalia Gimelshein}, \bibinfo{person}{Luca Antiga}, \bibinfo{person}{Alban Desmaison}, \bibinfo{person}{Andreas K{\"{o}}pf}, \bibinfo{person}{Edward~Z. Yang}, \bibinfo{person}{Zachary DeVito}, \bibinfo{person}{Martin Raison}, \bibinfo{person}{Alykhan Tejani}, \bibinfo{person}{Sasank Chilamkurthy}, \bibinfo{person}{Benoit Steiner}, \bibinfo{person}{Lu Fang}, \bibinfo{person}{Junjie Bai}, {and} \bibinfo{person}{Soumith Chintala}.} \bibinfo{year}{2019}\natexlab{}.
\newblock \showarticletitle{PyTorch: An Imperative Style, High-Performance Deep Learning Library}. In \bibinfo{booktitle}{\emph{Advances in Neural Information Processing Systems 32: Annual Conference on Neural Information Processing Systems 2019, NeurIPS 2019, December 8-14, 2019, Vancouver, BC, Canada}}. \bibinfo{pages}{8024--8035}.
\newblock
\urldef\tempurl%
\url{https://proceedings.neurips.cc/paper/2019/hash/bdbca288fee7f92f2bfa9f7012727740-Abstract.html}
\showURL{%
\tempurl}


\bibitem[Peng et~al\mbox{.}(2023)]%
        {RWKV2023}
\bibfield{author}{\bibinfo{person}{Bo Peng}, \bibinfo{person}{Eric Alcaide}, \bibinfo{person}{Quentin Anthony}, \bibinfo{person}{Alon Albalak}, \bibinfo{person}{Samuel Arcadinho}, \bibinfo{person}{Stella Biderman}, \bibinfo{person}{Huanqi Cao}, \bibinfo{person}{Xin Cheng}, \bibinfo{person}{Michael Chung}, \bibinfo{person}{Leon Derczynski}, \bibinfo{person}{Xingjian Du}, \bibinfo{person}{Matteo Grella}, \bibinfo{person}{Kranthi~Kiran GV}, \bibinfo{person}{Xuzheng He}, \bibinfo{person}{Haowen Hou}, \bibinfo{person}{Przemyslaw Kazienko}, \bibinfo{person}{Jan Kocon}, \bibinfo{person}{Jiaming Kong}, \bibinfo{person}{Bartlomiej Koptyra}, \bibinfo{person}{Hayden Lau}, \bibinfo{person}{Jiaju Lin}, \bibinfo{person}{Krishna Sri~Ipsit Mantri}, \bibinfo{person}{Ferdinand Mom}, \bibinfo{person}{Atsushi Saito}, \bibinfo{person}{Guangyu Song}, \bibinfo{person}{Xiangru Tang}, \bibinfo{person}{Johan~S. Wind}, \bibinfo{person}{Stanislaw Wozniak}, \bibinfo{person}{Zhenyuan Zhang}, \bibinfo{person}{Qinghua Zhou},
  \bibinfo{person}{Jian Zhu}, {and} \bibinfo{person}{Rui{-}Jie Zhu}.} \bibinfo{year}{2023}\natexlab{}.
\newblock \showarticletitle{{RWKV:} Reinventing RNNs for the Transformer Era}. In \bibinfo{booktitle}{\emph{Findings of the Association for Computational Linguistics: {EMNLP} 2023, Singapore, December 6-10, 2023}}. \bibinfo{publisher}{Association for Computational Linguistics}, \bibinfo{pages}{14048--14077}.
\newblock
\urldef\tempurl%
\url{https://doi.org/10.18653/V1/2023.FINDINGS-EMNLP.936}
\showDOI{\tempurl}


\bibitem[Qu et~al\mbox{.}(2024)]%
        {SSD4Rec2024}
\bibfield{author}{\bibinfo{person}{Haohao Qu}, \bibinfo{person}{Yifeng Zhang}, \bibinfo{person}{Liangbo Ning}, \bibinfo{person}{Wenqi Fan}, {and} \bibinfo{person}{Qing Li}.} \bibinfo{year}{2024}\natexlab{}.
\newblock \showarticletitle{SSD4Rec: {A} Structured State Space Duality Model for Efficient Sequential Recommendation}.
\newblock \bibinfo{journal}{\emph{CoRR}}  \bibinfo{volume}{abs/2409.01192} (\bibinfo{year}{2024}).
\newblock
\urldef\tempurl%
\url{https://doi.org/10.48550/ARXIV.2409.01192}
\showDOI{\tempurl}
\showeprint[arXiv]{2409.01192}


\bibitem[Quadrana et~al\mbox{.}(2018)]%
        {SRSurvey2018}
\bibfield{author}{\bibinfo{person}{Massimo Quadrana}, \bibinfo{person}{Paolo Cremonesi}, {and} \bibinfo{person}{Dietmar Jannach}.} \bibinfo{year}{2018}\natexlab{}.
\newblock \showarticletitle{Sequence-Aware Recommender Systems}.
\newblock \bibinfo{journal}{\emph{{ACM} Comput. Surv.}} \bibinfo{volume}{51}, \bibinfo{number}{4} (\bibinfo{year}{2018}), \bibinfo{pages}{66:1--66:36}.
\newblock
\urldef\tempurl%
\url{https://doi.org/10.1145/3190616}
\showDOI{\tempurl}


\bibitem[Rendle et~al\mbox{.}(2010)]%
        {FPMC2010}
\bibfield{author}{\bibinfo{person}{Steffen Rendle}, \bibinfo{person}{Christoph Freudenthaler}, {and} \bibinfo{person}{Lars Schmidt{-}Thieme}.} \bibinfo{year}{2010}\natexlab{}.
\newblock \showarticletitle{Factorizing personalized Markov chains for next-basket recommendation}. In \bibinfo{booktitle}{\emph{Proceedings of the 19th International Conference on World Wide Web, {WWW} 2010, Raleigh, North Carolina, USA, April 26-30, 2010}}. \bibinfo{publisher}{{ACM}}, \bibinfo{pages}{811--820}.
\newblock
\urldef\tempurl%
\url{https://doi.org/10.1145/1772690.1772773}
\showDOI{\tempurl}


\bibitem[Sarwar et~al\mbox{.}(2001)]%
        {CollaborativeFiltering2001}
\bibfield{author}{\bibinfo{person}{Badrul~Munir Sarwar}, \bibinfo{person}{George Karypis}, \bibinfo{person}{Joseph~A. Konstan}, {and} \bibinfo{person}{John Riedl}.} \bibinfo{year}{2001}\natexlab{}.
\newblock \showarticletitle{Item-based collaborative filtering recommendation algorithms}. In \bibinfo{booktitle}{\emph{Proceedings of the Tenth International World Wide Web Conference, {WWW} 10, Hong Kong, China, May 1-5, 2001}}. \bibinfo{publisher}{{ACM}}, \bibinfo{pages}{285--295}.
\newblock
\urldef\tempurl%
\url{https://doi.org/10.1145/371920.372071}
\showDOI{\tempurl}


\bibitem[Smith et~al\mbox{.}(2023)]%
        {S5-2023}
\bibfield{author}{\bibinfo{person}{Jimmy T.~H. Smith}, \bibinfo{person}{Andrew Warrington}, {and} \bibinfo{person}{Scott~W. Linderman}.} \bibinfo{year}{2023}\natexlab{}.
\newblock \showarticletitle{Simplified State Space Layers for Sequence Modeling}. In \bibinfo{booktitle}{\emph{The Eleventh International Conference on Learning Representations, {ICLR} 2023, Kigali, Rwanda, May 1-5, 2023}}. \bibinfo{publisher}{OpenReview.net}, \bibinfo{pages}{1--35}.
\newblock
\urldef\tempurl%
\url{https://openreview.net/forum?id=Ai8Hw3AXqks}
\showURL{%
\tempurl}


\bibitem[Srivastava et~al\mbox{.}(2014)]%
        {Dropout2014}
\bibfield{author}{\bibinfo{person}{Nitish Srivastava}, \bibinfo{person}{Geoffrey~E. Hinton}, \bibinfo{person}{Alex Krizhevsky}, \bibinfo{person}{Ilya Sutskever}, {and} \bibinfo{person}{Ruslan Salakhutdinov}.} \bibinfo{year}{2014}\natexlab{}.
\newblock \showarticletitle{Dropout: a simple way to prevent neural networks from overfitting}.
\newblock \bibinfo{journal}{\emph{J. Mach. Learn. Res.}} \bibinfo{volume}{15}, \bibinfo{number}{1} (\bibinfo{year}{2014}), \bibinfo{pages}{1929--1958}.
\newblock
\urldef\tempurl%
\url{https://doi.org/10.5555/2627435.2670313}
\showDOI{\tempurl}


\bibitem[Sun et~al\mbox{.}(2019)]%
        {BERT4Rec2019}
\bibfield{author}{\bibinfo{person}{Fei Sun}, \bibinfo{person}{Jun Liu}, \bibinfo{person}{Jian Wu}, \bibinfo{person}{Changhua Pei}, \bibinfo{person}{Xiao Lin}, \bibinfo{person}{Wenwu Ou}, {and} \bibinfo{person}{Peng Jiang}.} \bibinfo{year}{2019}\natexlab{}.
\newblock \showarticletitle{BERT4Rec: Sequential Recommendation with Bidirectional Encoder Representations from Transformer}. In \bibinfo{booktitle}{\emph{Proceedings of the 28th {ACM} International Conference on Information and Knowledge Management, {CIKM} 2019, Beijing, China, November 3-7, 2019}}. \bibinfo{publisher}{{ACM}}, \bibinfo{pages}{1441--1450}.
\newblock
\urldef\tempurl%
\url{https://doi.org/10.1145/3357384.3357895}
\showDOI{\tempurl}


\bibitem[Tang and Wang(2018)]%
        {TangW2018}
\bibfield{author}{\bibinfo{person}{Jiaxi Tang} {and} \bibinfo{person}{Ke Wang}.} \bibinfo{year}{2018}\natexlab{}.
\newblock \showarticletitle{Personalized Top-N Sequential Recommendation via Convolutional Sequence Embedding}. In \bibinfo{booktitle}{\emph{Proceedings of the Eleventh {ACM} International Conference on Web Search and Data Mining, {WSDM} 2018, Marina Del Rey, CA, USA, February 5-9, 2018}}. \bibinfo{publisher}{{ACM}}, \bibinfo{pages}{565--573}.
\newblock
\urldef\tempurl%
\url{https://doi.org/10.1145/3159652.3159656}
\showDOI{\tempurl}


\bibitem[Tran et~al\mbox{.}(2023)]%
        {MOJITO2023}
\bibfield{author}{\bibinfo{person}{Viet{-}Anh Tran}, \bibinfo{person}{Guillaume Salha{-}Galvan}, \bibinfo{person}{Bruno Sguerra}, {and} \bibinfo{person}{Romain Hennequin}.} \bibinfo{year}{2023}\natexlab{}.
\newblock \showarticletitle{Attention Mixtures for Time-Aware Sequential Recommendation}. In \bibinfo{booktitle}{\emph{Proceedings of the 46th International {ACM} {SIGIR} Conference on Research and Development in Information Retrieval, {SIGIR} 2023, Taipei, Taiwan, July 23-27, 2023}}. \bibinfo{publisher}{{ACM}}, \bibinfo{pages}{1821--1826}.
\newblock
\urldef\tempurl%
\url{https://doi.org/10.1145/3539618.3591951}
\showDOI{\tempurl}


\bibitem[Vaswani et~al\mbox{.}(2017)]%
        {Transformer2017}
\bibfield{author}{\bibinfo{person}{Ashish Vaswani}, \bibinfo{person}{Noam Shazeer}, \bibinfo{person}{Niki Parmar}, \bibinfo{person}{Jakob Uszkoreit}, \bibinfo{person}{Llion Jones}, \bibinfo{person}{Aidan~N. Gomez}, \bibinfo{person}{Lukasz Kaiser}, {and} \bibinfo{person}{Illia Polosukhin}.} \bibinfo{year}{2017}\natexlab{}.
\newblock \showarticletitle{Attention is All you Need}. In \bibinfo{booktitle}{\emph{Advances in Neural Information Processing Systems 30: Annual Conference on Neural Information Processing Systems 2017, December 4-9, 2017, Long Beach, CA, {USA}}}. \bibinfo{pages}{5998--6008}.
\newblock
\urldef\tempurl%
\url{https://proceedings.neurips.cc/paper/2017/hash/3f5ee243547dee91fbd053c1c4a845aa-Abstract.html}
\showURL{%
\tempurl}


\bibitem[Wu et~al\mbox{.}(2019)]%
        {SRGNN2019}
\bibfield{author}{\bibinfo{person}{Shu Wu}, \bibinfo{person}{Yuyuan Tang}, \bibinfo{person}{Yanqiao Zhu}, \bibinfo{person}{Liang Wang}, \bibinfo{person}{Xing Xie}, {and} \bibinfo{person}{Tieniu Tan}.} \bibinfo{year}{2019}\natexlab{}.
\newblock \showarticletitle{Session-Based Recommendation with Graph Neural Networks}. In \bibinfo{booktitle}{\emph{The Thirty-Third {AAAI} Conference on Artificial Intelligence, {AAAI} 2019, The Thirty-First Innovative Applications of Artificial Intelligence Conference, {IAAI} 2019, The Ninth {AAAI} Symposium on Educational Advances in Artificial Intelligence, {EAAI} 2019, Honolulu, Hawaii, USA, January 27 - February 1, 2019}}. \bibinfo{publisher}{{AAAI} Press}, \bibinfo{pages}{346--353}.
\newblock
\urldef\tempurl%
\url{https://doi.org/10.1609/AAAI.V33I01.3301346}
\showDOI{\tempurl}


\bibitem[Xu et~al\mbox{.}(2019)]%
        {LayerNorm2016}
\bibfield{author}{\bibinfo{person}{Jingjing Xu}, \bibinfo{person}{Xu Sun}, \bibinfo{person}{Zhiyuan Zhang}, \bibinfo{person}{Guangxiang Zhao}, {and} \bibinfo{person}{Junyang Lin}.} \bibinfo{year}{2019}\natexlab{}.
\newblock \showarticletitle{Understanding and Improving Layer Normalization}. In \bibinfo{booktitle}{\emph{Advances in Neural Information Processing Systems 32: Annual Conference on Neural Information Processing Systems 2019, NeurIPS 2019, December 8-14, 2019, Vancouver, BC, Canada}}. \bibinfo{pages}{4383--4393}.
\newblock
\urldef\tempurl%
\url{https://proceedings.neurips.cc/paper/2019/hash/2f4fe03d77724a7217006e5d16728874-Abstract.html}
\showURL{%
\tempurl}


\bibitem[Xu et~al\mbox{.}(2023)]%
        {Recbole-1.2.0}
\bibfield{author}{\bibinfo{person}{Lanling Xu}, \bibinfo{person}{Zhen Tian}, \bibinfo{person}{Gaowei Zhang}, \bibinfo{person}{Junjie Zhang}, \bibinfo{person}{Lei Wang}, \bibinfo{person}{Bowen Zheng}, \bibinfo{person}{Yifan Li}, \bibinfo{person}{Jiakai Tang}, \bibinfo{person}{Zeyu Zhang}, \bibinfo{person}{Yupeng Hou}, \bibinfo{person}{Xingyu Pan}, \bibinfo{person}{Wayne~Xin Zhao}, \bibinfo{person}{Xu Chen}, {and} \bibinfo{person}{Ji{-}Rong Wen}.} \bibinfo{year}{2023}\natexlab{}.
\newblock \showarticletitle{Towards a More User-Friendly and Easy-to-Use Benchmark Library for Recommender Systems}. In \bibinfo{booktitle}{\emph{Proceedings of the 46th International {ACM} {SIGIR} Conference on Research and Development in Information Retrieval, {SIGIR} 2023, Taipei, Taiwan, July 23-27, 2023}}. \bibinfo{publisher}{{ACM}}, \bibinfo{pages}{2837--2847}.
\newblock
\urldef\tempurl%
\url{https://doi.org/10.1145/3539618.3591889}
\showDOI{\tempurl}


\bibitem[Yue et~al\mbox{.}(2024)]%
        {LRURec2024}
\bibfield{author}{\bibinfo{person}{Zhenrui Yue}, \bibinfo{person}{Yueqi Wang}, \bibinfo{person}{Zhankui He}, \bibinfo{person}{Huimin Zeng}, \bibinfo{person}{Julian~J. McAuley}, {and} \bibinfo{person}{Dong Wang}.} \bibinfo{year}{2024}\natexlab{}.
\newblock \showarticletitle{Linear Recurrent Units for Sequential Recommendation}. In \bibinfo{booktitle}{\emph{Proceedings of the 17th {ACM} International Conference on Web Search and Data Mining, {WSDM} 2024, Merida, Mexico, March 4-8, 2024}}. \bibinfo{publisher}{{ACM}}, \bibinfo{pages}{930--938}.
\newblock
\urldef\tempurl%
\url{https://doi.org/10.1145/3616855.3635760}
\showDOI{\tempurl}


\bibitem[Zhang et~al\mbox{.}(2023)]%
        {TAT4SRec2023}
\bibfield{author}{\bibinfo{person}{Yihu Zhang}, \bibinfo{person}{Bo Yang}, \bibinfo{person}{Haodong Liu}, {and} \bibinfo{person}{Dongsheng Li}.} \bibinfo{year}{2023}\natexlab{}.
\newblock \showarticletitle{A time-aware self-attention based neural network model for sequential recommendation}.
\newblock \bibinfo{journal}{\emph{Appl. Soft Comput.}}  \bibinfo{volume}{133} (\bibinfo{year}{2023}), \bibinfo{pages}{109894}.
\newblock
\urldef\tempurl%
\url{https://doi.org/10.1016/J.ASOC.2022.109894}
\showDOI{\tempurl}


\bibitem[Zhang et~al\mbox{.}(2022)]%
        {AFMN2022}
\bibfield{author}{\bibinfo{person}{Yichi Zhang}, \bibinfo{person}{Guisheng Yin}, \bibinfo{person}{Hongbin Dong}, {and} \bibinfo{person}{Liguo Zhang}.} \bibinfo{year}{2022}\natexlab{}.
\newblock \showarticletitle{Attention-based Frequency-aware Multi-scale Network for Sequential Recommendation}.
\newblock \bibinfo{journal}{\emph{Appl. Soft Comput.}}  \bibinfo{volume}{127} (\bibinfo{year}{2022}), \bibinfo{pages}{109349}.
\newblock
\urldef\tempurl%
\url{https://doi.org/10.1016/J.ASOC.2022.109349}
\showDOI{\tempurl}


\bibitem[Zhu et~al\mbox{.}(2024)]%
        {VisionMamba2024}
\bibfield{author}{\bibinfo{person}{Lianghui Zhu}, \bibinfo{person}{Bencheng Liao}, \bibinfo{person}{Qian Zhang}, \bibinfo{person}{Xinlong Wang}, \bibinfo{person}{Wenyu Liu}, {and} \bibinfo{person}{Xinggang Wang}.} \bibinfo{year}{2024}\natexlab{}.
\newblock \showarticletitle{Vision Mamba: Efficient Visual Representation Learning with Bidirectional State Space Model}. In \bibinfo{booktitle}{\emph{Forty-first International Conference on Machine Learning, {ICML} 2024, Vienna, Austria, July 21-27, 2024}}. \bibinfo{publisher}{OpenReview.net}, \bibinfo{pages}{1--14}.
\newblock
\urldef\tempurl%
\url{https://openreview.net/forum?id=YbHCqn4qF4}
\showURL{%
\tempurl}


\bibitem[Zubic et~al\mbox{.}(2024)]%
        {SSMEC2024}
\bibfield{author}{\bibinfo{person}{Nikola Zubic}, \bibinfo{person}{Mathias Gehrig}, {and} \bibinfo{person}{Davide Scaramuzza}.} \bibinfo{year}{2024}\natexlab{}.
\newblock \showarticletitle{State Space Models for Event Cameras}. In \bibinfo{booktitle}{\emph{{IEEE/CVF} Conference on Computer Vision and Pattern Recognition, {CVPR} 2024, Seattle, WA, USA, June 16-22, 2024}}. \bibinfo{publisher}{{IEEE}}, \bibinfo{pages}{5819--5828}.
\newblock
\urldef\tempurl%
\url{https://doi.org/10.1109/CVPR52733.2024.00556}
\showDOI{\tempurl}


\end{thebibliography}

\end{document}